\documentclass[11pt]{article}

\usepackage[margin=1in]{geometry}
\usepackage{amsmath,amssymb}
\usepackage{graphicx}
\usepackage[utf8]{inputenc}
\usepackage[T1]{fontenc}
\usepackage[numbers,super,sort&compress]{natbib}
\usepackage[colorlinks=true,linkcolor=blue,citecolor=blue,urlcolor=blue]{hyperref}
\usepackage{setspace}
\usepackage{booktabs,array}

\graphicspath{{figs/}{.}}
\usepackage{placeins}
\newcolumntype{L}[1]{>{\raggedright\arraybackslash}p{#1}}
\newcommand{\OA}{\ensuremath{O_{A}}}
\newcommand{\OB}{\ensuremath{O_{B}}}
\newcommand{\icm}{cm$^{-1}$}

\begin{document}

\title{\textbf{Vibrational spectroscopy identifies the bond asymmetry of hexagonal diamond}}
\author{Li Zhu$^{1,\ast}$\\[6pt]
\normalsize $^{1}$Department of Physics, Rutgers University, Newark, New Jersey 07102, USA\\
\normalsize $^{\ast}$Corresponding author: li.zhu@rutgers.edu}
\date{}
\maketitle

\begin{abstract}
\noindent
Bulk hexagonal diamond has been synthesized by independent routes, but its structure remains
contested: the two recent refinements disagree on the sign of the difference between its two
bond lengths, by 238~m\AA, and both depart from a 2003 refinement. Here, we test the competing structures with first-principles lattice
dynamics. Relaxed hexagonal diamond has an axial bond
\emph{longer} than its three basal bonds by 24~m\AA, an eclipsed-bond effect that scales with
polytype hexagonality, and no realistic stress state, including the uniaxial formation stress,
inverts it. The bright $A_{1g}$ mode is the axial-bond stretch, shifting at
$\approx\!-2{,}100$~\icm~\AA$^{-1}$, while the E modes track the basal bonds, so the measured
splitting fixes the asymmetry at 0.41~m\AA{} per \icm{} independently of frequency calibration:
the spectrum can be inverted for the structure. Neither refined coordinate reproduces the
spectrum of its own sample. Inverting the phase-pure sample's spectrum gives
$\OB-\OA=24\pm3$~m\AA, and two determinations on separate samples give $+33\pm8$ and
$+60\pm45$~m\AA. The 1{,}529~\icm{} feature cannot be assigned to ideal 2H diamond, and the
local HRTEM observation remains open.
\end{abstract}

\section*{Introduction}

Hexagonal diamond (HD, lonsdaleite) has been pursued for six decades as the hexagonal counterpart
of cubic diamond (CD), with predicted mechanical properties rivalling or exceeding
it~\cite{Bundy1967,Frondel1967,Pan2009}. For most of that period it was accessible only as fine,
heavily faulted powders from shock or static compression, or as a minority component of
meteoritic carbon. Even its status as a distinct crystalline phase was contested: imaging and diffraction analyses argued that much of what had been reported
as lonsdaleite is better described as faulted and twinned cubic
diamond~\cite{Nemeth2014,Salzmann2015}, while shock-compression experiments followed its formation
from graphite in real time~\cite{Kraus2016,Turneaure2017}. Its defining structural feature is the
inequivalence of its bonds: each carbon forms three basal bonds (\OA), which link it into a
buckled honeycomb layer, and one axial bond (\OB) parallel to the hexagonal $c$ axis. In the orientation relationship observed in
every bulk synthesis, the axial bond descends from a bond \emph{within} a graphene sheet, and
the bond newly formed across the former gallery is one of the three basal bonds (Methods).
Whether \OB{} is shorter or longer than \OA, and
by how much, is the most elementary structural question about this material, and the literature
of the last two years has returned mutually exclusive answers to it. The question acquired urgency with the
recent wave of bulk-HD syntheses: from quasi-hydrostatic compression of single-crystal
graphite~\cite{Yang2025}, from heating of post-graphite phases~\cite{Chen2025}, and from direct
transformation of highly oriented pyrolytic graphite under predominantly uniaxial $c$-axis
stress~\cite{Yuan2025,Lai2026}.

Remarkably, the two structure determinations published since bulk synthesis was achieved
contradict each other. Yang
\textit{et al.}\cite{Yang2025} refine a \emph{short} axial bond ($\OB=1.510$ versus
$\OA=1.565$~\AA, from $z\approx0.0693$; their Extended Data Fig.~7) and match
their ultraviolet Raman spectrum with simulations at $\OB=1.44$~\AA. Lai \textit{et
al.}\cite{Lai2026}, for a millimetre-sized sample that is phase-pure by diffraction and
photoemission, refine a structure whose
axial bond is instead \emph{long} ($\OB=1.691$~\AA{} versus $\OA=1.508$~\AA, from
$z=0.0479(5)$; their Extended Data Table~1). The two refinements thus place the
bond asymmetry $\OB-\OA$ at $-55$ and $+183$~m\AA{} respectively, a 238~m\AA{} disagreement in a
material whose cubic polymorph has bond lengths known to sub-m\AA{} accuracy. Yang \textit{et al.}\cite{Yang2025}
additionally report a local high-resolution transmission-electron-microscopy (HRTEM) measurement of the same sign as their refinement but larger
magnitude, nominally $\OB-\OA=-80$~m\AA; we keep that observation distinct from the refinements
throughout, and return to it in the Discussion. The two reported structures have not been
reconciled.

Both also depart from an earlier determination that the recent literature has overlooked. Yoshiasa
\textit{et al.}\cite{Yoshiasa2003} refined hexagonal diamond synthesized in a Kawai-type
apparatus against a two-phase model of HD and stacking-faulted cubic diamond, and obtained
$c/a=1.668(6)$, an internal parameter $u=0.380(8)$ (equivalently $z=0.060(4)$), and an apical
(axial) bond \emph{longer} than the basal one, 1.59(4) versus
1.53(2)~\AA, giving $\OB-\OA=+60\pm45$~m\AA. Neither Yang \textit{et al.}\cite{Yang2025} nor Lai \textit{et al.}\cite{Lai2026}
cites this work. Measured on the common internal coordinate, and in units of the 2003 refinement's own
uncertainty ($\sigma=0.008$ on $u$), the 2003 value sits $0.7\sigma$ from
the relaxed first-principles structure reported below and $2.3\sigma$ and $3.0\sigma$ from the two
recent refinements respectively. We do not read that $0.7\sigma$ as a tight match: an
uncertainty of $\pm0.008$ is wide enough that agreement is easier to come by than
disagreement, and a loosely determined coordinate will sit close to many structures. The force of
that refinement lies elsewhere. Its axial ratio $c/a=1.668(6)$ is well determined ($5.8\sigma$
above the ideal 1.633) and by itself implies a long axial bond: at the ideal internal
coordinate, that cell already gives $\OB-\OA=+29$~m\AA. Its bond-length budget and stated
bond-strength ordering match ours, and the two recent refinements fall outside it in opposite
directions. The sign of its refined asymmetry, by contrast, is only a $1.3\sigma$ statement on
its own bond errors. The elementary question is therefore not open
in the sense of being untouched, but contested: three determinations exist, the oldest agrees with
theory, and the two newest disagree with it and with each other.

Here we show that the vibrational spectrum both groups measured settles this contradiction,
and that neither refined internal coordinate is consistent with the spectrum of its own sample. Density-functional-theory (DFT)
lattice dynamics establishes three facts. First, relaxed HD has a small, robust asymmetry of the \emph{opposite}
sign to Yang's determination: \OB{} exceeds \OA{} by 24~m\AA, in both a
generalized-gradient-approximation (GGA) and a dispersion-corrected meta-GGA functional, with an
identifiable local mechanism, the eclipsed
conformation of the axial bond. This is the sign, and close to the magnitude, that
Yoshiasa \textit{et al.}\cite{Yoshiasa2003} measured; theory here corroborates the oldest determination rather than
adjudicating alone. Second, the Raman spectrum is a quantitative gauge of the internal coordinate. The bright $A_{1g}$ zone-centre mode is essentially the stretch of the axial bond and moves at $\approx\!-2{,}100$~\icm~\AA$^{-1}$ with its length, whereas the two E modes track the
basal bonds. Because the two bond lengths enter the axial and basal modes with opposite sign, the
splitting between them fixes the asymmetry at 0.41~m\AA{} per \icm, while a common shift of all
frequencies leaves it unchanged. A measured spectrum can therefore be inverted for the structure
with a stated uncertainty, and its assignment is fixed by rules that follow from the same map. Whereas Yang \textit{et al.}\cite{Yang2025} used a one-variable frequency--bond relation to
assign a single band, the map developed here couples both bonds to all three modes, is shown to
be identifiable, and is validated on independent samples. Third, applying this gauge to the
published structures shows both Rietveld coordinates to be inconsistent with the spectra measured
on the same samples, within the ideal-crystal harmonic model, whereas the relaxed structure is
consistent with both at the level of the fitted component positions, and with a third,
independently measured lonsdaleite spectrum~\cite{Goryainov2018}. No mechanically viable,
diffraction-consistent $P6_3/mmc$ structure reproduces Yang's 1{,}529~\icm{} feature together
with their other two bands, leaving extrinsic sp$^2$ carbon (strained, disordered, or both)
as the leading candidate for its origin.

\section*{Results}

\subsection*{Relaxed hexagonal diamond has a longer axial bond}

\begin{figure}[tbp]\centering
\includegraphics[width=\textwidth]{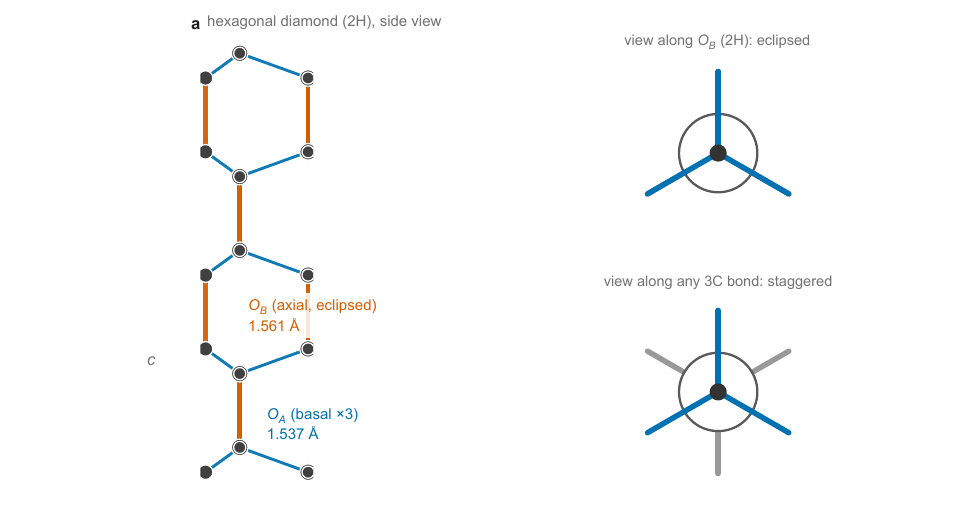}\\[2mm]
\includegraphics[width=\textwidth]{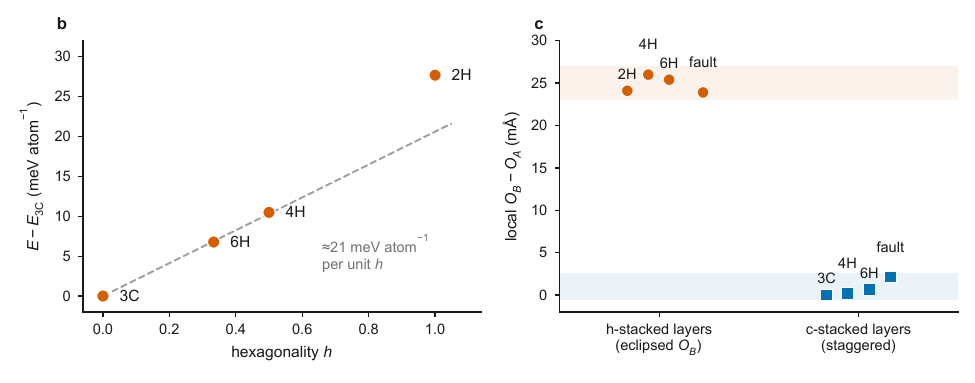}
\caption{\textbf{The eclipsed axial bond sets a local, transferable asymmetry.}
(a)~HD structure with basal \OA{} ($\times$3, blue) and axial \OB{} ($\times$1, orange);
the \OB{} bond is eclipsed, every CD bond staggered (Newman projections). (b)~Polytype energy versus hexagonality: $\approx$21~meV\,atom$^{-1}$ per unit
$h$, with 2H above the linear trend. (c)~Per-layer $\OB-\OA$ across 2H, 4H, 6H and a twinned
24-atom cell: hexagonally stacked layers $+24$--26~m\AA, cubically stacked 0--2~m\AA,
regardless of host.}
\label{fig:mech}
\end{figure}

Full relaxation of HD ($P6_3/mmc$, 4f Wyckoff; the 2H polytype in Ramsdell notation) with the
dispersion-corrected meta-GGA functional
r$^2$SCAN+rVV10~\cite{Furness2020,Peng2016,Ning2022} gives $a=2.5024$~\AA,
$c=4.1679$~\AA, $z=0.06277$, hence $\OA=1.537$~\AA{} ($\times$3) and $\OB=1.561$~\AA{}
($\times$1). The axial bond is thus \emph{longer} by 24.1~m\AA{} (Fig.~\ref{fig:mech}a).
The Perdew--Burke--Ernzerhof functional (PBE)~\cite{PerdewPBE} gives 23.4~m\AA{} with the same
sign. The asymmetry derives from $c/a=1.666$ exceeding the ideal
tetrahedral ratio 1.633; the bond-length budget is nearly conserved relative to cubic diamond
($3\OA+\OB=6.171$~\AA{} vs $4\times1.541=6.164$~\AA). Under pressure the asymmetry decreases only
weakly ($24.1\rightarrow20.5$~m\AA{} at 50~GPa), and bonding analysis assigns the shorter \OA{}
the larger bond strength (integrated crystal orbital Hamilton population, ICOHP: $-9.77$ vs
$-9.35$~eV at 0~GPa~\cite{Nelson2020}). (All four atoms occupy one
Wyckoff orbit, so atomic charges are symmetry-equivalent; the asymmetry resides in the bonds, not
in any ionic redistribution.)

Every one of these features was already reported experimentally\cite{Yoshiasa2003}: the same
sign of the asymmetry ($\OB-\OA=+60\pm45$~m\AA), and $c/a=1.668(6)$ against our 1.666. Their
mean C--C distance of 1.545~\AA, noted as ``very close to'' the cubic value, is our bond-budget
conservation. And their bonding conclusion, ``the enhanced covalency of basal bonds and the
reduced covalency of apical ones'' with ``the apical bond weaker than the basal bond'', is
precisely the ordering our ICOHP values reproduce. A 2003 powder refinement and a dispersion-corrected meta-GGA calculation thus
agree on the sign, the axial ratio, the bond budget and the bond-strength ordering. The axial ratio, the bond budget and the bond-strength ordering do not depend on the precision of
that refinement's internal coordinate, which is modest ($u=0.380(8)$). The sign is the
$1.3\sigma$ statement noted above, backed by the well-determined axial ratio. All four results
have been available, uncited, throughout the recent controversy.

Why does the lattice choose this distortion? The axial bond of 2H stacking is the
\emph{eclipsed} bond: the three back-bonds of its two atoms are in eclipsed conformation, whereas
every bond of cubic diamond is staggered (Fig.~\ref{fig:mech}a). Filled-orbital repulsion across
an eclipsed bond is expected to weaken and lengthen it; the same conformational preference explains the systematic
trends in internal parameter and axial ratio across the zinc-blende/wurtzite polytypes of the
tetrahedral semiconductors, carbon included~\cite{Yeh1992}. We test it at the crystal level
by a polytype construction (Fig.~\ref{fig:mech}b,c). Across 2H, 4H, 6H and a
24-atom twinned supercell, the local bond asymmetry of every layer follows its own stacking
character: hexagonally stacked (eclipsed) layers show $\OB-\OA=+24$ to $+26$~m\AA, cubically
stacked (staggered) layers 0 to $+2$~m\AA, irrespective of the host polytype
(Fig.~\ref{fig:mech}c). The total energy rises with hexagonality $h$ (the fraction of
hexagonally stacked layers) at $\approx$21~meV\,atom$^{-1}$ per unit $h$
($3\mathrm{C}\rightarrow6\mathrm{H}\rightarrow4\mathrm{H}$: 0, 6.8, 10.5~meV\,atom$^{-1}$). Pure
2H (27.7~meV\,atom$^{-1}$) lies above this linear trend, consistent with an additional repulsion
between adjacent eclipsed bonds once every axial bond is eclipsed. The asymmetry
of hexagonal diamond is thus a local, transferable property of hexagonal (eclipsed) stacking in
the polytypes examined, consistent with the conformational mechanism.

\subsection*{The asymmetry is small, strain-tunable, and erased at stacking faults}

\begin{figure}[tbp]\centering
\includegraphics[width=\textwidth]{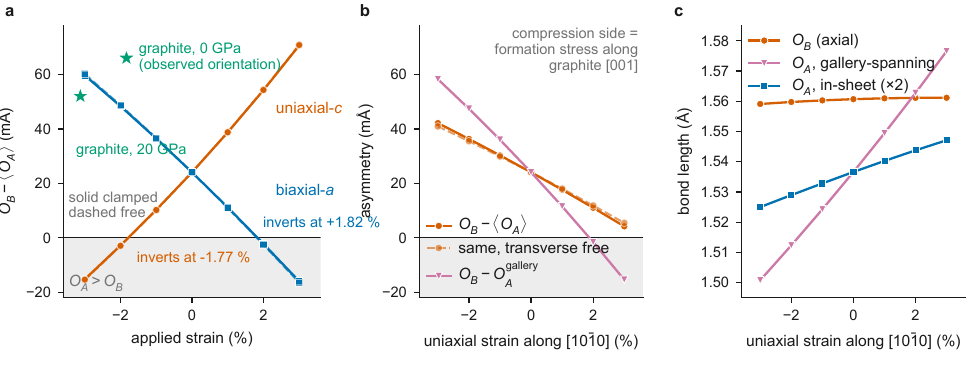}\\[2mm]
\includegraphics[width=120mm]{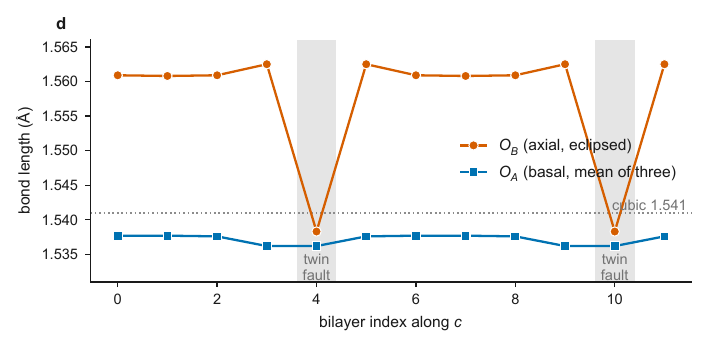}
\caption{\textbf{The asymmetry is delicate, but its sign is robust to the stresses of synthesis
and recovery.} (a)~$\OB-\langle\OA\rangle$ under uniaxial-$c$ and biaxial-$a$ strain with the
transverse dimensions clamped (solid) or free (dashed); the sign inverts at $-1.77\%$ and
$+1.82\%$ ($\approx$24~GPa of differential stress). Stars: HD coherent with graphite in the
observed orientation relationship (HD$[1\bar{2}10]\parallel$G$[1\bar{2}10]$,
HD$[0001]\parallel$G$[10\bar{1}0]$), plotted at the strain along $[1\bar{2}10]$. (b)~Uniaxial
strain along HD$[10\bar{1}0]$, the image of the graphite $[001]$ formation-stress axis: the mean
asymmetry (clamped, solid; transverse free, dashed) and the asymmetry against the
gallery-spanning basal bond; compression is the synthesis direction. (c)~The individual bonds
under the same strain. (d)~Local bonds across a twinned cell: cubic fault layers erase the
asymmetry toward the cubic value.}
\label{fig:strain}
\end{figure}

Although robust, the asymmetry is delicate on the scale of experimental perturbations
(Fig.~\ref{fig:strain}). Uniaxial compression along $c$ inverts its sign at
$\varepsilon_c\approx-1.8\%$, as does biaxial in-plane tension at $\varepsilon_a\approx+1.8\%$,
whether the transverse dimensions are clamped ($-1.79$ and $+1.84\%$) or left free to relax
($-1.77$ and $+1.82\%$). The Poisson coupling is weak because $C_{13}$ (8~GPa) is two orders of
magnitude below $C_{11}$ and $C_{33}$ (1{,}255 and 1{,}366~GPa). \OB{} couples almost exclusively
to $c$-strain ($d\OB/d\varepsilon_c\approx1.6$~\AA{} per unit strain) and \OA{} to $a$-strain. The stress scale matters: a strain of $\pm1.8\%$ corresponds to $\approx$24~GPa of differential
stress, so homogeneous residual strain cannot plausibly invert the asymmetry in a recovered
sample, though local stress at defects may do so in nanometric regions.

The stress under which bulk hexagonal diamond forms is not hydrostatic but predominantly uniaxial
along the graphite $c$ axis\cite{Lai2026,Yuan2025,Zhu2025JACS}, and in the orientation
relationship common to every bulk synthesis that axis becomes HD $[10\bar{1}0]$, an in-plane
direction of the product (Methods). We therefore computed the response to uniaxial strain along
$[10\bar{1}0]$ (Fig.~\ref{fig:strain}b,c). The strain lifts the hexagonal symmetry and separates
the three basal bonds into the one that spans the former graphite gallery, whose projection lies
along the strain axis, and the two that lie within the former sheet. Compression along this axis,
the direction of the synthesis stress, shortens the gallery-spanning bond most and
\emph{enlarges} the asymmetry, to $+30$, $+36$ and $+42$~m\AA{} at $-1$, $-2$ and $-3\%$.
Tension reduces it, but the mean asymmetry stays positive to $+3\%$ ($+4$~m\AA) and only the
single gallery-spanning bond overtakes \OB, beyond $\approx+2\%$. Clamped and transverse-free
responses agree to 1~m\AA, and uniaxial strain along $[2\bar{1}\bar{1}0]$ gives the same mean
asymmetry to within 1~m\AA{} (Supplementary Section~S10). Residual strain of the formation type
therefore cannot produce the reported inversion either. Once the lattice has relaxed, a bond's
length is set by the stress it is left in, not by the order in which it formed.

Coherency with residual graphite acts in the same direction. In the observed orientation, HD
$[1\bar{2}10]$ and $[0001]$ are locked to the graphite $a$ and $\sqrt{3}a$ periods. This
compresses the HD $a$ axis by 1.8\% and stretches its $c$ axis by 2.1\% at ambient pressure (by
3.1 and 0.7\% against the 20-GPa graphite lattice). With the third axis relaxed, the asymmetry
rises to $+66$~m\AA{} (ambient) and $+52$~m\AA{} (20~GPa). An explicit interface model shows that
the perturbation is short-ranged. The cells contain 48 atoms: six HD prismatic planes and six graphene layers in the observed
orientation, at either the HD or the graphite lateral lattice. Three starting geometries were
relaxed for each, including one in which the first sheet was pre-bonded to the HD surface in the
boat conformation. All relax to the same interface. The outermost pair of the diamond surface, an
axial pair, forms a 1.35~\AA{} carbon--carbon dimer, and the nearest graphene sheet settles
3.2~\AA{} away without a covalent bond to it. A sheet started at van der Waals distance does not
bond; the pre-bonded sheet stays bonded and becomes that reconstructed surface plane. The surface
plane loses most of its asymmetry, and the next sublevel inverts locally ($-21$~m\AA, through one
basal bond stretched to 1.63~\AA{} toward the reconstructed surface). The perturbation then decays
toward the interior value over the following sublevels: $+19$ to $+24$~m\AA{} at the HD lattice,
and $+60$ to $+66$~m\AA{} at the graphite lattice, whose interior value is the coherency value
(Supplementary Section~S11 and Supplementary Fig.~S4). In these models an incoherent graphite
contact thus perturbs a few atomic planes of diamond. Whether such interfaces contribute
measurably to a spectrum depends on their volume fraction, which the calculation does not
provide; the interface does offer one concrete candidate for locally inverted bonds. Cubic stacking faults
do not invert the asymmetry either; they locally erase it, the twin-boundary layers relaxing to
near-equal bonds (1.538/1.536~\AA, Fig.~\ref{fig:strain}d).

These results define what any measurement can see: the intrinsic bulk signal is $+24$~m\AA;
faulted regions contribute $\approx$0; interfaces perturb a few atomic planes; and sign inversion
requires tens of gigapascals of differential local stress, and even then amounts to only a few
m\AA. No computed realistic state, whether strained to $\pm3\%$ along any axis, coherent with or
in contact with graphite in the observed orientation, or stacking-faulted, produces the $-55$ to
$-80$~m\AA{} inversion reported by imaging and refinement\cite{Yang2025} beyond a single
interfacial plane.

\subsection*{The $A_{1g}$ mode is a gauge of the axial bond}

\begin{figure}[tbp]\centering
\includegraphics[width=\textwidth]{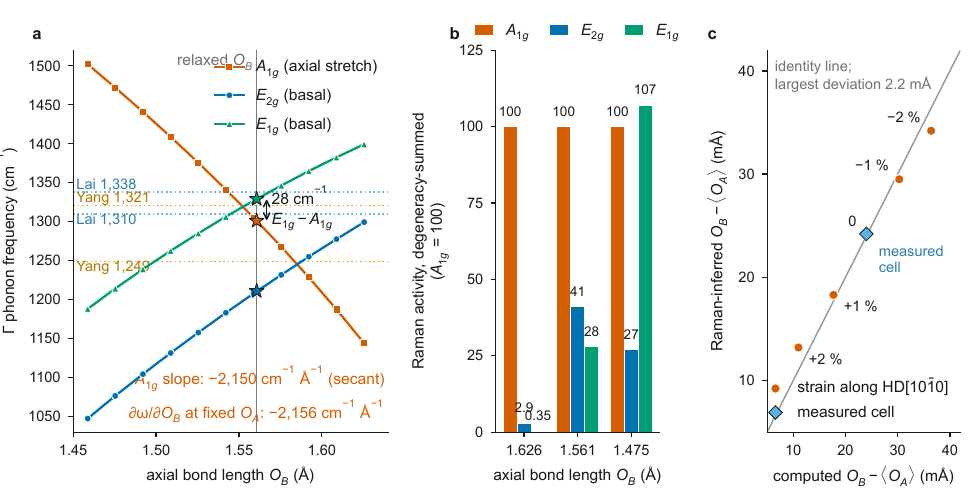}
\caption{\textbf{The $A_{1g}$ mode gauges the axial bond.} (a)~$\Gamma$ frequencies of the three
Raman-active modes versus \OB{} along the internal-coordinate path (r$^2$SCAN+rVV10; the PBE
control differs by a uniform $\sim$1.5\%; silent modes omitted, Source Data). Stars mark the
relaxed, unconstrained structure, whose $E_{1g}-A_{1g}$ splitting of 28~\icm{} is bracketed;
dotted lines are the measured bands\cite{Yang2025,Lai2026}. $A_{1g}$ disperses at
$\approx\!-2{,}100$~\icm~\AA$^{-1}$ (secant; $\partial\omega/\partial\OB|_{\OA}=-2{,}156$).
(b)~Degeneracy-summed Raman activities at three geometries (static, non-resonant, PBE;
$A_{1g}=100$). (c)~Validation on structures outside the calibration set: directly computed versus
Raman-inferred $\OB-\langle\OA\rangle$ for the clamped uniaxial HD$[10\bar{1}0]$ strain states
($-2$ to $+2\%$; exact phonons, mean of the split $E_{1g}$ pair read against $A_{1g}$) and for the
structure at the measured cell of the phase-pure sample; the line is the identity.}
\label{fig:gauge}
\end{figure}

Zone-center lattice dynamics along the fixed-cell internal-coordinate path (Methods) yields the
mode map of Fig.~\ref{fig:gauge}. The $A_{1g}$ mode is the axial-bond stretch, dispersing
from 1{,}144 to 1{,}502~\icm{} as \OB{} contracts from 1.626 to 1.459~\AA{} (secant slope
$-2{,}150$~\icm~\AA$^{-1}$; local slope $-2{,}600$ to $-1{,}800$ across the range). The
$E_{2g}$ and $E_{1g}$ modes track the basal bonds with opposite, weaker dependence. A
matched PBE calculation on identical structures reproduces this map with a uniform $\sim$1.5\%
softening and the same slopes, so the mapping is robust across the two functionals. We benchmark
the absolute frequency scale against experiment on known carbon modes computed with the identical
workflow: cubic-diamond $T_{2g}$ gives 1{,}325.5 vs measured 1{,}332~\icm~\cite{Solin1970}
($-0.5\%$) and the graphite G band gives 1{,}567 vs
$\approx$1{,}580~\icm~\cite{FerrariBasko2013} ($-0.8\%$). The absolute accuracy is thus
$\sim$1\%, consistent with the 12--26~\icm{} inter-functional spread used as the scale error
below.

The zone-center frequencies of \emph{relaxed} HD are themselves prior
art~\cite{Cui2015,FloresLivas2012,Denisov2011,ElMendili2022}. Our values of 1{,}211, 1{,}301 and
1{,}329~\icm{} agree
closely with the ideal-HD frequencies tabulated in Fig.~3a of ref.~\citenum{Yang2025} (1{,}218,
1{,}300 and 1{,}332~\icm, attributed there to refs.~\citenum{Cui2015,FloresLivas2012}), which were
obtained with different functionals, and with independent predictions of the same manifold
(1{,}221/1{,}280/1{,}338~\icm~\cite{Denisov2011}; 1{,}207/1{,}307/1{,}330~\icm~\cite{ElMendili2022}).
Raman spectra of stacking-disordered diamond have likewise been computed and connected to the
phonon density of states~\cite{Murri2019,Denisov2011}. That independent calculations concur on the equilibrium spectrum to within
$\sim$10~\icm{} is a premise of what follows rather than a result of it. This has a direct
consequence for the measurements: an anomalously short axial bond inferred from a spectrum
cannot be attributed to a deficiency of one functional, because the functionals agree on the
ideal structure. The equilibrium frequencies accordingly serve
here as calibration. What is new is the sensitivity map that follows, and the inversion of measured
spectra for the bond asymmetry that it makes possible.

To establish that the gauge is a property of the bond rather than of the scan path, we computed a
fixed-bond response surface: structures with \OA{} held at 1.58~\AA{} while \OB{} varies (each
with the remaining cell parameter energy-minimized), and conversely \OA{} varied at fixed \OB.
The resulting partial derivatives separate cleanly. For the $A_{1g}$ mode,
$\partial\omega/\partial\OB|_{\OA}=-2{,}156$~\icm~\AA$^{-1}$ with a weak cross-sensitivity
$\partial\omega/\partial\OA|_{\OB}\approx-300$; the E modes behave oppositely
($\partial\omega(E_{1g})/\partial\OA|_{\OB}=-2{,}725$;
$\partial\omega(E_{2g})/\partial\OA|_{\OB}=-3{,}025$; \OB{} cross-terms $\approx+300$). The
$A_{1g}$ is thus an axial-bond gauge and the E modes are basal-bond gauges. This holds robustly
across constructions. The $A_{1g}$ \OB-sensitivity is $-2{,}000\pm150$~\icm~\AA$^{-1}$
whether evaluated along the fixed-cell path, the fixed-\OA{} energy-minimized path, or the
fixed-\OA{} fixed-$c$ closure (these constructions relax different cell parameters, so their
$\sim$7\% spread bounds the closure dependence). The bond force constants confirm this. The
longitudinal force constants extracted from the equilibrium force-constant matrix are
14.2~eV\,\AA$^{-2}$ for \OA{} and 12.1~eV\,\AA$^{-2}$ for \OB, against 13.9~eV\,\AA$^{-2}$ for
the cubic-diamond bond. The axial bond is therefore longer, weaker by bond population, and 15\%
softer than the basal bonds, which are in turn 17\% stiffer than it (Supplementary
Section~S12).

The gauge was then tested on structures that played no part in its calibration. These were the
clamped uniaxial strain states along HD$[10\bar{1}0]$, which break the hexagonal symmetry and
split each E mode, and the structure at the measured lattice constants of the phase-pure
sample. For each,
the $\Gamma$ phonons were computed exactly and the mean of the split $E_{1g}$ pair was read
against $A_{1g}$ through the calibrated 0.41~m\AA{} per \icm. The inferred asymmetry follows the
directly computed one from $+11$ to $+36$~m\AA{} with a largest deviation of 2.2~m\AA{}
(Fig.~\ref{fig:gauge}c; Supplementary Section~S10). The linearized map therefore recovers a
changed asymmetry, not only the equilibrium value it was built around.

We now invert the measured spectrum. The phase-pure sample\cite{Lai2026} yields one
asymmetric envelope fitted with \emph{two} components, near 1{,}310 and 1{,}338~\icm{} (their
Supplementary Fig.~4); the frequently quoted 1{,}227~\icm{} value belongs to their
\emph{calculated} spectrum (the $E_{2g}$ position), not to an independently resolved experimental
peak. Passing the two measured components through the DFT-calibrated sensitivity map yields a
structure of $\OA=1.533$~\AA{} and $\OB=1.557$~\AA, within 5~m\AA{} of the relaxed prediction.
It gives $\OB-\OA=24\pm3$~m\AA{} (an estimated propagated uncertainty, Supplementary
Section~S1) \emph{for that sample}; the scope of this
interval and its behaviour on other material are discussed below. This interval is narrow
because the asymmetry is set by the \emph{splitting} of the two components, not their absolute
positions.
Shifting all frequencies together by 1~\icm{} (a common additive offset $\delta$, our nuisance
model, bounded to $|\delta|\lesssim15$~\icm{} by the CD/graphite benchmarks) moves the inferred
$\OB-\OA$ by only 0.005~m\AA, whereas changing the splitting by 1~\icm{} moves it by 0.41~m\AA;
the measured splitting, 28~\icm, coincides with the calculated $E_{1g}-A_{1g}$ of 28. It is likewise insensitive to a multiplicative scale factor,
which changes a 28~\icm{} splitting by only $\sim$0.3~\icm{} at the benchmarked 1\% level.
Effects that move the components together cancel for the same reason, whether they come from
calibration, residual hydrostatic strain, phonon confinement or temperature. Their computed
differential parts are small. Confinement in crystallites down to 3~nm shifts the $E_{1g}$ and
$A_{1g}$ components by at most 0.15~\icm{} relative to each other. The quasi-harmonic
(thermal-expansion) shift at 300~K differs between them by 0.01~\icm{} (mode Gr\"uneisen
parameters 0.96 and 1.02). Together they contribute below 0.1~m\AA{} (Supplementary
Section~S13). The explicit phonon--phonon shift, about $-1$ to $-2$~\icm{} for the diamond line
between 0 and 300~K\cite{Lang1999}, has not been computed mode by mode here. The budget allows its
differential part to be as large as that whole shift ($\pm0.8$~m\AA); a differential shift of
5~\icm{} would move the inferred asymmetry by 2~m\AA. Resonant excitation reweights intensities but
does not move first-order zone-centre frequencies, which are ground-state properties; the
excitation-dependent bands of sp$^2$ carbon are defect-activated double-resonance features, a
different mechanism\cite{FerrariBasko2013}. Reweighting can, however, shift the fitted position
of an overlapping component, which is part of the component-fit uncertainty the interval
carries. The
inference is also robust to the choice of lattice constants: enforcing the sample's measured cell
($a=2.5179$, $c=4.1828$~\AA) instead of the calculated one shifts both inferred bonds together by
$+12$~m\AA{} (to 1.545/1.569~\AA) and leaves their difference at 24~m\AA; the common shift is
absorbed by the offset $\delta$, enlarged to $+29$~\icm{} in quantitative agreement with the
mode-Gr\"uneisen softening expected from the 1.6\% cell-volume difference between the calculated
and measured lattices, and confirmed by exact phonons at the measured-cell structure
(Supplementary Section~S1). The recovered samples are also well inside the validity window of the gauge. The lattice
constants published for the three bulk samples ($a=2.518$--$2.519$, $c=4.178$--$4.183$~\AA) and
for the 2003 material ($a=2.508$, $c=4.183$~\AA) agree with each other to 0.5\% and with the
calculated cell to 0.6\% in $a$ and 0.4\% in $c$. The average residual differential strain of the
recovered material is therefore at most a few tenths of a per cent, an order of magnitude below
the $\pm1.8\%$ that inverts the asymmetry, and corresponds to at most 2--3~m\AA{} in
Fig.~\ref{fig:strain}. Local strains around defects are not bounded by these averages, but the
validation above shows that they bias the inference little. Over $\pm2\%$ of uniaxial in-plane
strain the inverted value tracks the true, strain-changed asymmetry to 2~m\AA, so residual strain
changes what the gauge measures rather than how accurately it measures it. A residual uniaxial
in-plane stress would in addition split each E mode by $\approx$20~\icm{} per per cent of strain
(Supplementary Section~S10). Lai \textit{et al.}\ fit the E component as a single line, which
indicates that no splitting was resolved at the component width. The published data do not fix
the splitting that would have been detectable, so we treat this as an indication rather than a
bound. The quoted interval combines component-fit uncertainty, the functional spread of the
reference, the 7\% closure spread of the sensitivities, the computed confinement and
quasi-harmonic bounds and the anharmonic allowance. It excludes zero, Yang's $-55$~m\AA{} and
Lai's $+183$~m\AA{} by wide margins.

The assignment itself is testable, and all six ordered assignments of the two components to the
three modes were inverted (Supplementary Section~S1). Reading the components as
$A_{1g}$/$E_{2g}$ requires $\OA=1.495$~\AA{} ($\sim$3\% in-plane compression, $\approx$42~GPa
in-plane stress, and $a$ incompatible with the measured lattice) and predicts a strong unobserved
$E_{1g}$ band near 1{,}443~\icm. One alternative is not excluded by mechanics: reading 1{,}310 as
$E_{1g}$ and 1{,}338 as $A_{1g}$ inverts to a near-symmetric structure ($\OB-\OA\approx+1$~m\AA)
close to equilibrium, at low energy and stress. It fails against the spectra instead, on three
counts. Exact phonons and Raman tensors computed at that structure show that its third mode
falls at 1{,}191~\icm, in-window and unobserved. At this near-equal-bond geometry the three band
activities are nearly equal ($98{:}100{:}103$ for $E_{2g}{:}A_{1g}{:}E_{1g}$), so the missing
1{,}191~\icm{} band is predicted to be as strong as the two observed components. The same near-equal activities also give no account of the
strongly asymmetric measured envelope, whose dominant 1{,}310~\icm{} component the preferred
assignment reproduces as the bright $A_{1g}$. And on the one sample where all three modes are
resolved~\cite{Goryainov2018}, the E-splitting bound of Supplementary Section~S4 forces
$A_{1g}$ to be the middle, most intense band (see below), the ordering of every published
calculation including Lai's own~\cite{Lai2026}. Within the HD spectral models examined, the $A_{1g}$/$E_{1g}$ assignment is therefore the one
consistent with the observed intensity ordering, the third-mode constraint and the resolved
third-sample spectrum. The quoted interval is conditional on it and on the visibility of the
$E_{1g}$ component in the reported geometry (full assignment matrix in Supplementary
Information).

The upper component also lies close to the first-order line of cubic diamond, so a minority CD
contribution must be considered. The component sits 5--6~\icm{} above the
ambient CD frequency of 1{,}332--1{,}333~\icm~\cite{Solin1970}, and the CD line shifts by
2.9~\icm~GPa$^{-1}$ under pressure\cite{Boppart1985}, so a compressive stress of about 2~GPa would
bring it to the fitted position. The published characterization supports an HD-dominated specimen:
phase-pure by diffraction and photoemission according to its authors, with no CD or graphite
signal identified in the Raman response~\cite{Lai2026}. A single sp$^3$ photoemission peak does
not by itself distinguish the two sp$^3$ phases, however. The available spectral data do not
allow an independent quantitative comparison of mixed-phase line-shape models. We therefore adopt the $A_{1g}$/$E_{1g}$ assignment within the HD model. Lai
\textit{et al.}'s own reading of the envelope, three modes near 1{,}227, 1{,}310 and
1{,}338~\icm~\cite{Lai2026}, coincides with that assignment's prediction
(1{,}221/1{,}310/1{,}338~\icm).

Crucially, the $A_{1g}$ mode also dominates the Raman intensity near equilibrium, a pattern
established experimentally by Goryainov \textit{et al.}\cite{Goryainov2018}, who isolated all
three Raman-active modes of lonsdaleite and reported the $A_{1g}$ band at 1{,}305~\icm{} as ``the most intense band
in the experimental Raman spectrum'', and independently by a natural diamond--lonsdaleite sample
whose most intense band, at 1{,}307~\icm, is likewise attributed to
$A_{1g}$~\cite{ElMendili2022}. Finite-field
Raman tensors give degeneracy-summed relative band activities
$A_{1g}:E_{2g}:E_{1g}=100:41:28$ at the relaxed geometry (static, non-resonant, PBE level;
Methods). The $A_{1g}$ band is overwhelmingly dominant at long bonds ($100:3:0.4$ at
$\OB=1.63$~\AA) and remains the brightest band down through equilibrium; at the shortest bond
sampled ($\OB=1.475$~\AA) the $E_{1g}$ band becomes comparable ($\approx$107\% of $A_{1g}$). The
predicted spectrum of relaxed HD is therefore a single asymmetric band: a dominant $A_{1g}$ at
1{,}301~\icm{} overlapped by the weaker $E_{1g}$ at 1{,}329~\icm, with a smaller $E_{2g}$ component
at 1{,}211~\icm. This is consistent with the reported spectrum of the phase-pure sample of Lai
\textit{et al.}, one asymmetric envelope whose two fitted components (1{,}310 and 1{,}338~\icm)
lie within 9~\icm{} of the predicted $A_{1g}$ and $E_{1g}$~\cite{Lai2026}. The predicted $E_{2g}$
at 1{,}211~\icm{} is not separately resolved in their two-component fit; it could overlap the
unresolved low-frequency side of the envelope, and its visibility depends on the scattering
geometry of the oriented specimen. In backscattering along $c$, $E_{1g}$ is symmetry-forbidden
while $A_{1g}$ and $E_{2g}$ are allowed; in edge-on geometries all three appear. The reported
instrument is a 355-nm \emph{microconfocal} platform~\cite{Lai2026}. Focusing and collection
through a microscope objective span a cone of propagation directions, and no polarization
analysis is described, so the strict backscattering selection rules are relaxed and $E_{1g}$ can
appear. Its actual weight depends on the numerical aperture, the crystal orientation and the
polarization configuration, none of which is reported. The $E_{1g}$ band can therefore be
visible; its intensity is not verified. Polarization-resolved measurements would settle the assignment
directly. Their own simulated spectrum, computed independently with the same functional
class, shows the identical three-mode $A_{1g}$-dominant pattern, including the $E_{2g}$ near
1{,}227~\icm. One assignment made on frequency alone needs correction. The
$B_{1g}$ mode near 1{,}273~\icm,
an apparent match to the 1{,}249~\icm{} band, is forbidden in first-order Raman scattering for
ideal $P6_3/mmc$ HD by the selection rules for the diamond polytypes~\cite{Spear1990} (disorder or
local symmetry breaking could weakly activate it) and cannot carry a principal assignment.

The gauge can be tested against material of entirely different provenance. Goryainov \textit{et al.}\cite{Goryainov2018}
extracted a lonsdaleite spectrum from Popigai impact diamonds by subtracting the spectra of two
samples of differing hexagonality, and resolved all three Raman-active modes: $E_{2g}=1{,}244$,
$A_{1g}=1{,}305$ and $E_{1g}=1{,}356$~\icm. That assignment is not merely adopted from those
authors; the mode map forces it. Of the three ways to pair two of their bands as the E modes,
only $1{,}244/1{,}356$ (splitting 112~\icm) falls inside the 107--131~\icm{} window accessible
to the E-mode splitting anywhere on the sampled manifold (Supplementary Section~S4), fixing
$A_{1g}=1{,}305$: the most intense band, the ordering our activities predict, and a measured
E-splitting that lands mid-window. Their $A_{1g}$ lies 4~\icm{} from our relaxed
prediction of 1{,}301~\icm, an independent confirmation of the gauge's absolute position on a
third sample, from a natural impact rather than a synthesis. Their $A_{1g}$--$E_{1g}$ splitting,
however, is 51~\icm{} against our 28, which through the same sensitivity
($0.41$~m\AA{} per \icm) maps to $\OB-\OA\approx33$~m\AA{} rather than 24. The discrepancy is
within the
uncertainty those authors themselves attach to the measurement. Their bands are
``highly broadened (50--60~\icm)'' and ``slightly ($\sim$10~\icm) shifted toward low energies due
to lonsdaleite imperfection'', and the spectrum is a difference of two spectra of a mixture
containing between 0 and 42\% lonsdaleite. Propagating a component-position uncertainty of a
quarter to a third of the band width makes the splitting uncertain by $\approx$20~\icm, hence
$\OB-\OA=33\pm8$~m\AA, an interval that overlaps the phase-pure one. We therefore quote
$24\pm3$~m\AA{} as the value for the phase-pure sample\cite{Lai2026}, where the components are
narrow and separately fitted. Three experimental determinations on independent samples are now
available: $+24\pm3$ (Raman, phase-pure), $+33\pm8$ (Raman, impact diamond~\cite{Goryainov2018})
and $+60\pm45$~m\AA{} (diffraction~\cite{Yoshiasa2003}). They agree in sign and overlap, and none
of them is compatible with either recent refinement. The three uncertainties are of different
kinds and not on a common statistical footing. The first two are our propagated estimates
(Supplementary Section~S1), and the third is the refinement's own bond error; the two Raman
values share our calculated sensitivity map, which the diffraction value does not.

\subsection*{The measured spectra discriminate the proposed structures}

\begin{figure}[tbp]\centering
\includegraphics[width=\textwidth]{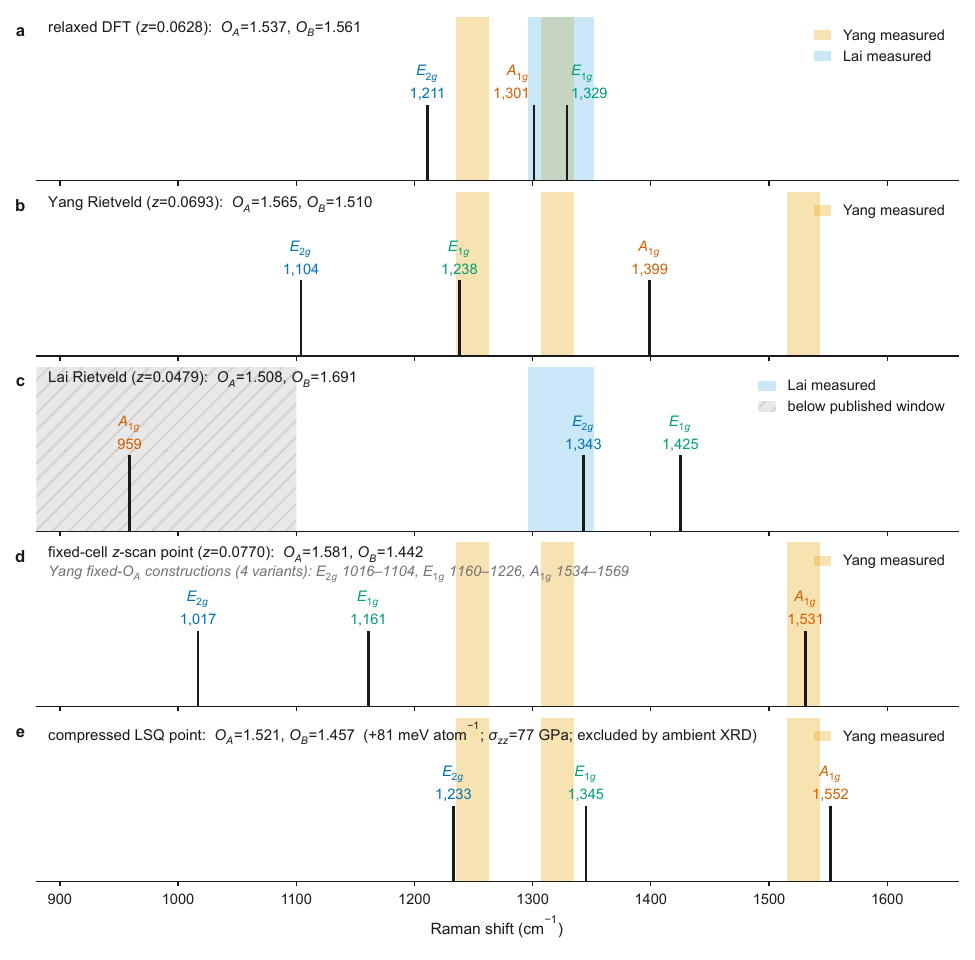}
\caption{\textbf{Measured band positions discriminate the proposed structures.} Predicted
Raman-active mode positions (equal-height sticks, identical convention for every model) of the
five structures of Table~\ref{tab:decisive} against the measured band positions of the same
samples (shaded; fitted components for ref.~\citenum{Lai2026}). Hatched: below the published
measurement window of ref.~\citenum{Lai2026}. Shaded regions are visual guides of fixed width and
do not represent confidence intervals; calculated activities are shown separately
(Fig.~\ref{fig:gauge}b) and weighted by intensity in Supplementary Information.
Panel~d shows the fixed-cell $z$-scan realization of the short-\OB{} assignment (the four
fixed-\OA{} Yang-construction variants are annotated); panel~e shows the compressed
least-squares structure, which matches all three Yang bands but requires
$\sigma_{zz}=77$~GPa confinement and is excluded by the ambient lattice constants of Yang's own
XRD.}
\label{fig:decisive}
\end{figure}

The gauge can now be applied to every published structure (Table~\ref{tab:decisive};
Fig.~\ref{fig:decisive}). For each structure we compute the $\Gamma$ phonons directly (with
residual-force subtraction; Methods), so all entries are calculated, not interpolated.

\begin{table}[tb]
\centering
\caption{\textbf{Predicted Raman-active modes (\icm) of candidate and published HD structures
versus the spectrum measured on the same sample.} Measured bands for reference: ref.~\citenum{Yang2025}
1{,}249/1{,}321/1{,}529~\icm; ref.~\citenum{Lai2026} 1{,}310/1{,}338~\icm{} (two fitted components of
one envelope). The ``sample'' column gives which measurement each row is judged against.
Theoretical scale error $\le$1--2\% (15--30~\icm), calibrated by the PBE control and the
CD/graphite benchmarks. Bold marks the $A_{1g}$ band. ``Activities'': degeneracy-summed Raman activities
$E_{2g}{:}A_{1g}{:}E_{1g}$ from exact finite-field tensors at each structure (static, PBE;
$A_{1g}=100$). LSQ: the exact-DFT check point in the least-squares island of
Fig.~\ref{fig:map}. Verdicts summarize pattern-level
comparisons (see text for measurement-window and coincidence caveats).}
\label{tab:decisive}
\footnotesize
\setlength{\tabcolsep}{2.0pt}
\begin{tabular}{L{2.1cm}cccccccL{2.3cm}}
\toprule
structure & $z$ & \OA/\OB{} (\AA) & $E_{2g}$ & $A_{1g}$ & $E_{1g}$ & activities & sample & verdict \\
\midrule
relaxed DFT & 0.0628 & 1.537/1.561 & 1{,}211 & \textbf{1{,}301} & 1{,}329 & 41:100:28 & both & best account of 1{,}249--1{,}321 envelope; 1{,}529 unassigned \\
Rietveld, ref.~\citenum{Yang2025} & 0.0693 & 1.565/1.510 & 1{,}104 & \textbf{1{,}399} & 1{,}238 & 129:100:317 & Yang & pattern not consistent \\
Rietveld, ref.~\citenum{Lai2026} & 0.0479 & 1.508/1.691 & 1{,}343 & \textbf{959} & 1{,}425 & 0.1:100:1.2 & Lai & in-window E pair 82~\icm{} apart vs 28 measured; not consistent \\
compressed LSQ point & 0.0674 & 1.521/1.457 & 1{,}233 & \textbf{1{,}552} & 1{,}345 & 120:100:323 & Yang & matches, but excluded by XRD + stress \\
$z$-scan point, $\OB=1.442$~\AA & 0.0770 & 1.581/1.442 & 1{,}017 & \textbf{1{,}531} & 1{,}161 & 8:100:51 & Yang & high band only (1{,}529 $\checkmark$; 1{,}249/1{,}321 $\times$) \\
Yang fixed-\OA{} constr.\ (4 variants) & n/a & 1.56--1.58/1.44 & 1{,}016--1{,}104 & \textbf{1{,}534--1{,}569} & 1{,}160--1{,}226 & -- & Yang & high band only (1{,}529 $\checkmark$; 1{,}249/1{,}321 $\times$) \\
\bottomrule
\end{tabular}
\end{table}

The relaxed structure gives the best account of the common Raman envelope of \emph{both}
samples. For Lai \textit{et al.} it matches to within 9~\icm{} on each of the two fitted
components. For Yang \textit{et al.} the dominant envelope agrees to 8--20~\icm, while the weak
1{,}249~\icm{} component leaves a 38~\icm{} residual against $E_{2g}$, the largest surviving
discrepancy (and plausibly strain- or fit-related in the twinned material; Discussion). Neither
experimental Rietveld structure is consistent with the spectrum measured on its own sample at the
pattern level. Each contains a near-coincidence (Yang-Rietveld $E_{1g}$ 1{,}238 vs the
1{,}249 band; Lai-Rietveld $E_{2g}$ 1{,}343 vs the 1{,}338 component), yet both leave measured
components unexplained and predict additional modes with no observed counterpart,
78--145~\icm{} from the nearest measured band. (The Lai-Rietveld $A_{1g}$ discrepancy of 351~\icm{} on symmetry assignment
involves a mode below the published window and is not counted as an observed absence.)
Discrepancies of this size exceed the 12--26~\icm{} functional spread by factors of 3--10,
and anharmonic, thermal or resonance effects are unlikely to account for them under the reported
ambient conditions. The two refinements moreover contradict each other ($z=0.0693$ vs 0.0479,
bracketing the relaxed 0.0628). Lai's refinements carry signatures of an ill-conditioned fit:
preferred-orientation texture indices of 5.5 (HD-only) and 4.0 (mixed model), and atomic
displacement parameters $U_{\rm iso}=0.032$--$0.040$~\AA$^2$, roughly an order of magnitude above
typical diamond values. The refinement details of ref.~\citenum{Yang2025} are more sparsely reported.
Refining a light-atom internal coordinate from a textured, stacking-faulted powder is a
recognized difficulty, not a matter of care. Preferred orientation acts most strongly on the
$00l$ reflections that carry the $z$ information\cite{Dollase1986}, and the disagreement between
the two refinements reflects the limits of the method for this material rather than the care of
either group.
Lai \textit{et al.}'s own Supplementary Table~1 shows the sensitivity: refining the same pattern
with an HD-only model or with an HD/CD mixture moves $z$ from 0.0479(5) to 0.0504(5). That change
of $25\times10^{-4}$, comparable to the quantity at issue, comes from the phase model alone.
A diffraction sensitivity analysis of the published Lai pattern (Supplementary Information) makes
the ill-conditioning quantitative. When scale, displacement parameter and a one-parameter texture proxy are allowed to covary at
each $z$, the pattern shows little discriminatory power across the entire range
$z=0.040$--$0.080$, with the classic degeneracy signatures: $B_{\rm iso}$ pinned at
zero, the texture parameter tracking $z$, and the minimum position depending on the covariance
treatment. The root cause is the severe preferred orientation, which acts most strongly on the
$z$-sensitive $00l$ reflections. In every covariance treatment examined, the fit at
$z\approx0.063$--$0.070$ is at least as good as at the refined 0.0479, so the published pattern
does not discriminate against the DFT-relaxed coordinate. (The original refinement used a more
flexible
spherical-harmonic texture model, so our one-parameter proxy provides a sensitivity analysis
rather than a reproduction of that refinement.)

Two of these comparisons need care: the measurement window in one case, coincidental
single-mode matches in the other. The published
Lai spectrum spans $\approx$1{,}100--1{,}700~\icm, so the Lai-Rietveld structure's strongest
predicted band ($A_{1g}$ at 959~\icm) lies \emph{below} the published range and its
non-observation is not by itself evidence. Exact Raman tensors computed at that coordinate
(Table~\ref{tab:decisive}) show, moreover, that it would place most of its scattering strength in
that sub-window band: the in-window $E_{2g}$ (1{,}343~\icm) and $E_{1g}$ (1{,}425~\icm) carry
0.1\% and 1\% of the $A_{1g}$ activity. The in-window spectrum it predicts is thus weak relative
to a band that was not measured. Relative activities set no absolute detection limit, so this is
supporting rather than decisive. The decisive comparison is one of positions. The two in-window
modes of that structure are 82~\icm{} apart, whereas the two measured components are 28~\icm{}
apart, and no common offset reconciles the two patterns; the near-coincidence of its $E_{2g}$
with the 1{,}338 component does not change that. The mixed HD/CD refinement of
ref.~\citenum{Lai2026} ($z=0.0504$) behaves identically: $A_{1g}=1{,}032$, $E_{2g}=1{,}325$,
$E_{1g}=1{,}414$~\icm{} (an in-window separation of 89~\icm, with 0.6\% and 1.5\% of the activity
in the window), so neither refinement is consistent with the spectrum (Supplementary Table~S3).
At the Yang coordinate the exact tensors show the $A_{1g}$ mode close to its activity
minimum (its $\alpha_{zz}$ changes sign between the relaxed and the short-bond geometries). The
strongest predicted band is therefore the $E_{1g}$ at 1{,}238~\icm, near the observed 1{,}249
band, with $E_{2g}$ at 1{,}104~\icm{} second (129\% of $A_{1g}$) and $A_{1g}$ at 1{,}399~\icm{}
weakest. That
near-coincidence is therefore not a weak accident, and the pattern still does not match: the
strongest measured band, at 1{,}321~\icm, has no counterpart, the predicted second band at
1{,}104~\icm{} is absent, and 1{,}529~\icm{} is unexplained. (Because of the sign change of
$\alpha_{zz}$, the activity ordering at these coordinates cannot be interpolated along the
internal-coordinate scan; interpolation would wrongly suggest $A_{1g}$ dominance at the Yang
coordinate. The tensors must be computed at each structure, as done here.)

The remaining rows of Table~\ref{tab:decisive} address the Raman-based structure assignment of
Yang \textit{et al.}\cite{Yang2025}. At $\OB=1.442$~\AA{} the $A_{1g}$ mode sits at 1{,}531~\icm, reproducing
their assignment of the 1{,}529~\icm{} band exactly. But the \emph{same} structure places $E_{2g}$
and $E_{1g}$ at 1{,}017 and 1{,}161~\icm, missing the other two measured bands by 160--230~\icm.
Could \emph{some} homogeneous 2H structure reproduce all three bands at once? A least-squares
search of the two-bond manifold (Fig.~\ref{fig:map}; search construction and uniqueness checks in
Supplementary Information) identifies one candidate region, doubly compressed, with its
linearized minimum at $(\OA,\OB)=(1.514,1.451~\text{\AA})$. Exact phonons computed at a point
inside it, $(1.521,1.457~\text{\AA})$, give 1{,}233/1{,}552/1{,}345~\icm{} and match Yang's three
bands to 16--24~\icm{} (Table~\ref{tab:decisive}). That residual is not random but dominated by a single
discrepancy, the predicted E-mode splitting of 112~\icm{} against the 72~\icm{} observed, which
no structure on the manifold can close (Supplementary Section~S4). This structure is nonetheless
unphysical for a recovered ambient sample, on two independent grounds. First, it lies
81~meV\,atom$^{-1}$ above the relaxed structure and needs a confining stress of $\sigma_{xx}=22$~GPa
in-plane and $\sigma_{zz}=77$~GPa axial (calculated stress tensor); even under that stress it is
not internally stationary (residual atomic forces of $\approx$2~eV\,\AA$^{-1}$, Supplementary
Table~S2), so external stress alone could not hold it. Second, and decisively, its
lattice constants ($a=2.464$, $c=3.990$~\AA) are 2.2\% and 4.5\% \emph{smaller} than the ambient
values from Yang's own X-ray diffraction (XRD) measurement ($a=2.5182$, $c=4.1780$~\AA), so their diffraction excludes the only
structure whose Raman matches all three of their bands. The two probes therefore close the argument
jointly: the Raman spectrum excludes the refined coordinate, and the measured lattice constants
exclude the compressed Raman-matching family. No mechanically viable, diffraction-consistent
homogeneous 2H structure reproduces (1{,}249, 1{,}321, 1{,}529). Under the measured ambient lattice
constraint, the two lower-frequency bands select the relaxed structural region, and the
1{,}529~\icm{} feature is left unassigned.

\begin{figure}[tbp]\centering
\includegraphics[width=\textwidth]{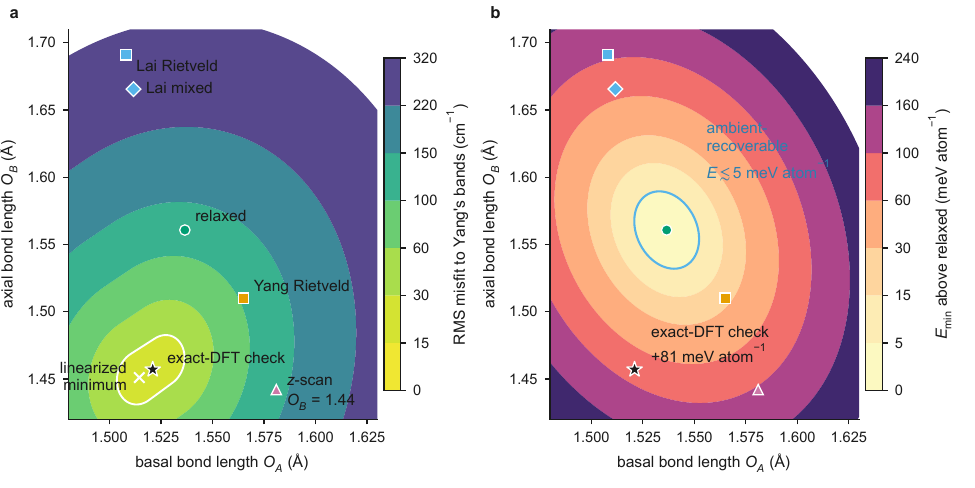}
\caption{\textbf{Joint-probe structural inference on the $(\OA,\OB)$ manifold.}
(a)~Root-mean-square (RMS) misfit of the linearized mode map to Yang's three bands
(symmetry-assigned), minimized over a common offset bounded to the benchmarked $\pm$15~\icm{}
(Supplementary Section~S4, Eq.~5); the white contour is 30~\icm. The only low-misfit region is
doubly compressed; its linearized minimum ($\times$) is at $(1.514,1.451)$~\AA, and exact DFT
phonons were computed at the starred point inside it, $(1.521,1.457)$~\AA, whose linearized misfit
is within 1~\icm{} of the minimum. (b)~Minimum energy above the relaxed structure at fixed
$(\OA,\OB)$, $E_{\min}$, from a quadratic fit to the six computed structures whose remaining cell
parameter is energy-minimized (Supplementary Section~S4; rms residual 2~meV\,atom$^{-1}$): the
low-misfit region lies $\gtrsim$80~meV\,atom$^{-1}$ above equilibrium, requires tens-of-GPa
confining stress ($\sigma_{zz}=77$~GPa at the starred point), and its lattice constants are
excluded by the ambient XRD of the same sample. Spectroscopy admits a mathematical solution away
from equilibrium, but mechanics and diffraction exclude it, leaving only the relaxed region.
}
\label{fig:map}
\end{figure}

We also tested Yang's own construction directly. Their Methods compute frequencies for ``various
\OA{} and \OB{} lengths while maintaining the hexagonal unit cell parameters and symmetry''; the
published text fixes \OA{} at 1.58~\AA, while the precursor preprint states 1.56~\AA. We therefore
built the fixed-\OA{} construction exhaustively at $\OB=1.44$~\AA: with \OA{} at either value, and
the remaining cell freedom either energy-minimized or fixed to their experimental $c$ (four
variants). All four reproduce their $A_{1g}$ ($1{,}534$--$1{,}569$~\icm{} vs their 1{,}529) but
none approaches their E-mode values: $E_{2g}=1{,}016$--$1{,}104$ and $E_{1g}=1{,}160$--$1{,}226$~\icm,
short of the published 1{,}249 and 1{,}321~\icm{} by 145--233 and 95--161~\icm{} respectively (the
$\OA=1.56\rightarrow1.58$ shift itself follows the sensitivity map, $+55$--$60$~\icm). Indeed,
across the entire fixed-\OA{} path ($\OB=1.44$--$1.60$~\AA) the E modes never approach those
values. The discrepancy with their Fig.~3b thus persists on every variant of their stated
construction, for both functionals: we could not reproduce the reported E-mode trends under the
stated structural constraints.

\subsection*{Candidate origins of the high-frequency band}

The unassigned high-frequency band has a natural extrinsic candidate. The G band of graphitic
carbon softens under in-plane tension at $-58.5$~\icm{} per \% (r$^2$SCAN+rVV10; unstrained
1{,}567~\icm{} vs experimental $\approx$1{,}580~\icm~\cite{FerrariBasko2013}), reaching
1{,}529~\icm{} at $\approx$0.7--0.9\% tension (0.65\% against the calculated intercept, 0.87\%
after aligning to the experimental G position; Supplementary Fig.~S3). Our computed slope sits in
the middle of the measured and calculated consensus for biaxial strain: $-63$~\icm/\% from the
Gr\"uneisen parameter and $-58$ from density-functional theory\cite{Mohiuddin2009},
$-62.6\pm4$ from the purpose-built equibiaxial measurement\cite{Androulidakis2015}, and
$-57$ from strained graphene bubbles~\cite{Zabel2012}. The last of these is close to a direct
precedent for the present assignment. Zabel \textit{et al.}\cite{Zabel2012} measured the graphene G peak at
1{,}525~\icm{} at the centre of a bubble under $\approx$1\% biaxial strain, against 1{,}598~\icm{}
on the flat substrate. A band within a few \icm{} of 1{,}529 is thus an established consequence of
about one per cent of biaxial tension in sp$^2$ carbon. This is consistent with minority sp$^2$
carbon partially stretched toward registry with the surrounding diamond matrix at twins or grain
boundaries.

A competing extrinsic explanation must be acknowledged. Along the amorphisation trajectory of
sp$^2$ carbon the G peak falls from $\approx$1{,}600 to $\approx$1{,}510~\icm{} with no strain at
all~\cite{FerrariRobertson2000}, so disorder alone can place a band near 1{,}529~\icm. The two
mechanisms are distinguishable, but not by position: strained crystalline sp$^2$ should retain a
narrow G band with a weak D peak and little dispersion with excitation energy, whereas the
amorphisation route requires a broad G band, a strong and strongly dispersing D peak, and a
falling $I(\mathrm{D})/I(\mathrm{G})$ in the nanocrystalline-to-amorphous regime. Neither
linewidths nor multi-wavelength data are reported for the 1{,}529~\icm{} band, so we cannot
separate them here, and we present strain and disorder as the two viable extrinsic candidates
rather than asserting the first. Both remain preferable to a first-order mode of homogeneous ideal
2H HD, which the cross-sample systematics below exclude on independent grounds. The attribution must respect the constraints reported by Yang \textit{et al.}\cite{Yang2025}: no G band at the
unstrained position, and $\approx$3\% surface-sensitive sp$^2$ by electron energy-loss
spectroscopy (attributed by those authors to focused-ion-beam damage). The carrier would therefore be a minority component, plausibly visible
through the strong ultraviolet resonance of sp$^2$ carbon. The alternative is nanoscale HD locally
compressed to $\OB\approx1.44$~\AA, more than 100~meV\,atom$^{-1}$ above the relaxed structure
along the constrained path. It cannot be excluded outright, but it predicts a companion feature that the spectrum lacks.
Exact Raman tensors at $\OB=1.442$~\AA{} give $100{:}8{:}51$ for $A_{1g}{:}E_{2g}{:}E_{1g}$
(Table~\ref{tab:decisive}). Any fraction of the sample that produced 1{,}529~\icm{} by this mechanism would therefore also
produce an $E_{1g}$ band near 1{,}161~\icm{} at about half that intensity, regardless of how
much relaxed material contributes at 1{,}321~\icm. Nothing is reported near 1{,}161~\icm. This companion-band test is independent of the absolute frequency scale
and of the functional, with the caveat that the activities are static and non-resonant. The
relaxed structure, whose dominant $A_{1g}$ (1{,}301~\icm) overlapped by $E_{1g}$ (1{,}329~\icm)
coincides with the strongest measured band, needs no such companion.
The two candidates are experimentally distinguishable. The HD $A_{1g}$ tensor
is dominated by $\alpha_{zz}$ ($\alpha_{zz}/\alpha_{xx}\approx-5$ at equilibrium) whereas the G
band is in-plane, so polarization-resolved ultraviolet Raman on oriented material would decide.

The cross-sample systematics strongly favour a sample-dependent extrinsic origin. Across the four
published bulk-HD Raman measurements, a lower-frequency response in the
$\approx$1{,}250--1{,}340~\icm{} region, the fingerprint of the relaxed structure, recurs in every
sample. Yang \textit{et al.} report bands at 1{,}249 and 1{,}321~\icm\cite{Yang2025}, Yuan \textit{et al.}
$E_{2g}=1{,}246$ and $A_{1g}=1{,}312$~\icm~\cite{Yuan2025}, and Lai \textit{et al.} two fitted
components at 1{,}310 and 1{,}338~\icm\cite{Lai2026}. Chen \textit{et al.} report a broad
$\approx$1{,}315~\icm{} band~\cite{Chen2025}, resolved by those authors into three Gaussian
components assigned, by reference to calculated spectra, to the $E_{2g}$/$A_{1g}$/$E_{1g}$
manifold. The fitted positions are not reported; the positions computed here,
1{,}211/1{,}301/1{,}329~\icm, span that band and supply what the assignment lacked. The
high-frequency response, by contrast, varies from sample to sample: 1{,}529~\icm{} in the
triple-twinned material\cite{Yang2025}; a broad $\approx$1{,}550~\icm{} band that its authors
attribute to residual graphitic phases\cite{Chen2025}; bands at 1{,}459 and 1{,}600~\icm{} in
ref.~\citenum{Yuan2025}; and \emph{no}
high band in the phase-pure, 54-arcsec-rocking-curve sample\cite{Lai2026}. Yuan \textit{et al.}\cite{Yuan2025}
themselves assign their 1{,}459 and 1{,}600~\icm{} bands to ``a trace amount
of sp$^2$ carbon remaining in grain boundaries'', consistent with a weak $\pi^\ast$ feature in
their EELS. Chen \textit{et al.}\cite{Chen2025} assign their 1{,}550~\icm{} band to ``the
small amount of residual graphitic phases'' within the disordered-carbon framework of
ref.~\citenum{FerrariRobertson2000}. Two of the four groups thus read their own high-frequency
band as extrinsic sp$^2$ carbon. This response varies by over 140~\icm{} and vanishes entirely in the cleanest sample,
which is difficult to reconcile with a first-order mode of homogeneous ideal 2H HD. A minority
sp$^2$ component with sample-dependent strain and abundance does exactly this, as two of the
synthesis groups have already concluded for their own material. The one-phonon route can moreover be closed
entirely: the phonon density of states of relaxed HD, computed from the same force constants on
a $24\times24\times16$ wavevector mesh, terminates at 1{,}329~\icm{} (Supplementary
Section~S8), 200~\icm{} below the observed band. Disorder-activated scattering samples this
density of states~\cite{Denisov2011}, and Raman spectra of stacking-disordered diamond evolve
within the same manifold~\cite{Murri2019}, so no one-phonon sp$^3$ channel, zone-centre or not,
reaches 1{,}529~\icm{} at the relaxed structure; fault layers themselves relax toward the cubic
geometry (Fig.~\ref{fig:strain}d). The intrinsic candidates that remain are multiphonon
scattering, which we have not computed, and the locally compressed regions addressed above; any
candidate must also survive the cross-sample systematics.

\section*{Discussion}

Three statements summarize the resolution. (i)~The intrinsic structure of hexagonal diamond has a
\emph{long} axial bond, $\OB-\OA=+24$~m\AA, an eclipsed-bond effect that is local to
hexagonal stacking, transferable across the polytypes tested, strain-tunable, and erased at cubic
faults. The sign, the axial ratio, the bond budget and the bond-strength ordering all reproduce
the 2003 refinement\cite{Yoshiasa2003}. (ii)~The two recent diffraction refinements of
bulk HD are inconsistent with each other, with the Raman spectra of their own samples (within
the ideal-crystal harmonic model), and with the earlier refinement, at $2.3\sigma$ and
$3.0\sigma$; their $z$-coordinates should be used with caution
pending re-refinement. (iii)~The vibrational spectrum, specifically the bright $A_{1g}$ axial
stretch, discriminates the proposed structures far more sharply than the available diffraction,
and it supports the relaxed structure in both existing bulk samples and in impact
diamond~\cite{Goryainov2018}. Taken together these remove the asymmetry from the domain in which
theory arbitrates between experiments: the calculation now agrees with the oldest measurement and
with every measured spectrum, and disagrees only with the two most recent refined coordinates.

The remaining experimental tension is the local HRTEM bond-length measurement of
Yang \textit{et al.}\cite{Yang2025} ($\OA=1.58\pm0.05$, $\OB=1.50\pm0.08$~\AA{} over a
$0.8\times1.2$~nm$^2$ region). No computed realistic state reproduces its nominal
$-80$~m\AA{} inversion. We note that the quoted uncertainties are per-bond spreads over 18--20 bonds; that
the raw in-image averages differ from the calibrated values; and that a single nanometric region
of a twinned, recovered sample is precisely where local strain and fault proximity make the
asymmetry least representative. Two further considerations apply. Projected distances in a
high-resolution image shorten anisotropically under a small specimen tilt. The geometric
foreshortening alone is small, about 6~m\AA{} at $5^\circ$ for an in-plane bond, so tilt is a
possible systematic to be documented rather than an established explanation of an 80~m\AA{}
difference. The interface model above also shows that one diamond sublevel next to the
reconstructed surface at an incoherent graphite contact is locally inverted, which supplies one
concrete candidate for such a region. A
statistical HRTEM survey across many regions, with the tilt documented, and polarization-resolved
ultraviolet Raman are the discriminating measurements we propose.

Beyond clarifying the controversy, the eclipsed-bond picture makes testable predictions. For the
polytypes tested (3C, 6H, 4H, 2H), the mean bond asymmetry scales linearly with
hexagonality, at $\approx$25~m\AA{} per unit $h$, and we predict this to extend to other
stacking sequences. The $A_{1g}$
frequency of faulted material should shift with stacking content, suggesting that Raman spectra
report the average stacking order within the optical probe volume. Differential stress tunes the
asymmetry and inverts it only near 24~GPa, whereas hydrostatic compression preserves its sign to
at least 50~GPa. A residual uniaxial in-plane stress would show up as a splitting of each E mode by $\approx$20~\icm{} per per cent of strain (Supplementary Section~S10).

The construction may extend beyond carbon, given material-specific calibration. Every wurtzite-type
structure has one internal parameter that sets an axial and a basal bond. Whether an axial and a
basal mode separate as cleanly as here must be established for each material, by computing its
sensitivity matrix and checking identifiability. In polar wurtzites the LO--TO splitting and the
mixing of the $E_1$ and $E_2$ modes add terms the carbon map does not contain. The natural targets
are non-polar materials whose internal coordinate is hard to refine from powder diffraction
because of texture, stacking faults or light atoms: the higher diamond polytypes (4H, 6H) and
hexagonal silicon and germanium. Wurtzite boron nitride, whose internal parameter was refined
together with lonsdaleite in the 2003 study\cite{Yoshiasa2003}, would need the polar
corrections. For faulted material the
map predicts that the axial-mode frequency shifts with stacking content, so the spectrum reports
the average stacking order and, through the E-mode splitting, the stress state within the probe
volume.

Finally, the eclipsed-bond picture connects the structure to the transformation that makes it.
In the orientation relationship observed for every bulk synthesis, each graphene sheet becomes a
puckered $(10\bar{1}0)$ plane of hexagonal diamond, the axial bond descends from a bond within
the sheet, and the bond formed across the former gallery is a basal bond. Whether a buckling
sheet adopts the boat conformation, which makes that in-sheet bond eclipsed and yields hexagonal
diamond, or the chair conformation, which makes it staggered and yields cubic diamond, is the
polytype-selecting step identified in molecular-dynamics studies of the graphite
transformation\cite{Chen2024JACSAu,Zhu2025JACS}. That step is favoured by uniaxial stress along
the graphite $[001]$ axis at moderate temperature\cite{Lai2026,Luo2022}. The conformational
variable that selects the polytype is therefore the same one that sets the bond asymmetry. The
delicacy of the eclipsed bond quantified here, inversion at $\pm1.8\%$ of differential strain but
no inversion under compression along the formation axis, constrains how much residual strain a
nucleation model may leave in the product.

\section*{Methods}

\textbf{Density-functional calculations.} VASP~6.4.2~\cite{Kresse1996,KresseCMS1996}, projector
augmented-wave (PAW) potentials~\cite{Blochl1994,KresseJoubert1999}, plane-wave cutoff 800~eV, EDIFF~$=10^{-8}$~eV.
Primary functional r$^2$SCAN+rVV10~\cite{Ning2022}, combining
r$^2$SCAN~\cite{Furness2020,FurnessErratum2020} with the rVV10 dispersion
correction~\cite{Peng2016} at the refitted BPARAM~$=11.95$~\cite{Ning2022}; control
calculations with PBE~\cite{PerdewPBE}. $\Gamma$-centred k-meshes: $19\times19\times11$ (4-atom
HD cell); commensurate meshes for supercells ($7\times7\times5$ for $3\times3\times2$).
Structural relaxations to forces $<10^{-3}$~eV\,\AA$^{-1}$. Polytype and fault supercells were
built from a tetrahedral-bilayer generator validated by recovering the relaxed 2H and 3C
structures; 4H (ABCB), 6H (ABCACB) and a 24-atom twinned 2H cell were relaxed with variable cell
shape. Throughout, \OA{} denotes the three basal bonds and \OB{} the axial bond. The basal
bonds are symmetry-equivalent as long as the $P6_3/mmc$ symmetry is retained (the fixed-bond
constructions and the internal-coordinate scan keep it); where a strain lowers the symmetry they
are reported individually.

\textbf{Bond correspondence under the graphite--diamond orientation relationship.} Every bulk
synthesis reports G$(0001)\parallel$HD$(10\bar{1}0)$, G$[10\bar{1}0]\parallel$HD$[0001]$ and
G$[1\bar{2}10]\parallel$HD$[1\bar{2}10]$\cite{Yang2025,Lai2026,Chen2025,Yuan2025}. Building the
relaxed HD structure in that frame shows that the bonds parallel to G$[10\bar{1}0]$ are the axial
bonds (1.561~\AA). The bonds spanning between adjacent $(10\bar{1}0)$ prismatic planes, which are
the bonds formed across the former graphite galleries, are basal bonds (1.537~\AA). Each graphene
sheet becomes a puckered prismatic plane containing one axial and two basal bonds per atom. The former graphite $[001]$ axis becomes HD$[10\bar{1}0]$.

\textbf{Stress states.} Cartesian strain tensors were applied to the relaxed cell (uniaxial
along $c$, equibiaxial in the basal plane, uniaxial along $[10\bar{1}0]$ and along
$[2\bar{1}\bar{1}0]$; $-3$ to $+3\%$) and the internal coordinates relaxed at fixed cell. For the
transverse-free states the transverse strains were set from the computed elastic tensor and
verified to leave every transverse stress below 0.3~GPa. Coherency states lock the HD periods
along $[1\bar{2}10]$ and $[0001]$ to the graphite $a$ and $\sqrt3 a$ periods (r$^2$SCAN+rVV10
graphite, 0 and 20~GPa) and minimize the energy over the third axis. Interface cells contain six
HD $(10\bar{1}0)$ prismatic planes and six graphene layers (48 atoms) stacked along the common axis
HD$[10\bar{1}0]\parallel$G$[0001]$, with the lateral cell fixed at either the HD or the graphite
lattice. Three starting geometries were relaxed for each: graphene 1.8 and 3.3~\AA{} above the
diamond surface, and a sheet pre-bonded at the gallery spacing with boat buckling. Relaxation was
run first at 600~eV and then at 800~eV, with the cell length adjusted to zero stress along the
stacking axis; residual forces are below 0.012~eV\,\AA$^{-1}$.

\textbf{Lattice dynamics.} Finite-displacement phonons
(phonopy~2.48~\cite{Togo2015,Togo2023}),
$3\times3\times2$ supercells (72 atoms), two symmetry-inequivalent displacements per structure.
For constrained (non-stationary) structures, namely the internal-coordinate scan and the
experimentally refined geometries, residual forces of the undisplaced supercell were computed and
subtracted; we verified that this reproduces the unconstrained equilibrium result to $<$4~\icm. Mode symmetries from the irreducible-representation decomposition at $\Gamma$. The
internal-coordinate scan spans $z=0.055$--$0.077$ at the relaxed lattice vectors (the $z=0.077$
point computed as a separate constrained structure); experimentally refined structures were
computed at their published lattice constants and coordinates (ref.~\citenum{Yang2025} Extended Data
Fig.~7: $a=2.5182$, $c=4.1780$~\AA, $z=0.0693$; ref.~\citenum{Lai2026} Extended Data Table~1:
$a=2.5179$, $c=4.1828$~\AA, $z=0.0479$). The fixed-bond response surface holds $(\OA,\OB)$ fixed
and resolves the remaining cell degree of freedom by energy minimization (documented closure; an
alternative closure at the experimental $c$ changes $A_{1g}$ by $<$20~\icm). $\Gamma$ phonons at
the $[10\bar{1}0]$-strained cells use the same protocol with the displacements required by the
lowered symmetry; mode Gr\"uneisen parameters were obtained from $\Gamma$ phonons at the cells
relaxed at 0, 5 and 10~GPa. Longitudinal bond force constants are $-\hat{u}^{\rm T}\Phi_{ij}\hat{u}$
for the symmetrized force-constant block $\Phi_{ij}$ of each bonded pair and its unit vector
$\hat{u}$; the phonon-confinement bound follows the Richter--Campbell model with the branch
dispersions from the same force constants (Supplementary Section~S13).

\textbf{Raman activities.} Finite-difference polarizability derivatives: mode-projected
displacements of $\pm0.02$~\AA{} along $\Gamma$ eigenvectors, macroscopic dielectric tensors from
density-functional perturbation theory (PBE), powder band activities
$45\bar\alpha'^2+7\bar\gamma'^2$ summed over degenerate partners. The tensors were computed at
the relaxed structure, at two scan geometries and at every refined or candidate structure of
Table~\ref{tab:decisive}, each with the eigenvectors of its own residual-force-subtracted force
constants. Activities are static and non-resonant; both experiments used ultraviolet excitation
(325/355~nm), where resonance effects can reweight intensities. Frequencies are therefore the
primary evidence and intensities supportive. The $A_{1g}$-dominant pattern is corroborated by the published simulated spectrum\cite{Lai2026} and by the reported component positions.

\textbf{Inverting a measured spectrum for the internal coordinate.} The procedure used here
applies to a Raman spectrum of hexagonal diamond in which the $A_{1g}$ component and at least one
E component are resolved and identified; two E components alone fix \OA{} but not the asymmetry.
(i)~\emph{Model.} About the relaxed structure ($\OA^0=1.5366$, $\OB^0=1.5607$~\AA;
$\omega^0=1{,}211.0$, $1{,}301.2$, $1{,}329.2$~\icm{} for $E_{2g}$, $A_{1g}$, $E_{1g}$) the mode
positions are $\omega_i=\omega_i^0+J_i\cdot(\Delta\OA,\Delta\OB)+\delta$. The sensitivity rows
$J_{E_{2g}}=(-3{,}025,+300)$, $J_{A_{1g}}=(-300,-2{,}156)$ and $J_{E_{1g}}=(-2{,}725,+300)$~\icm~\AA$^{-1}$
come from the fixed-bond response surface. The common additive offset $\delta$ absorbs the
frequency calibration, the functional scale error (benchmarked at $|\delta|\le15$~\icm{} at the
calculated cell) and the mode-Gr\"uneisen softening of a measured cell (Supplementary
Section~S1). (ii)~\emph{Assignment.} The E-mode splitting depends on \OA{} alone,
$\omega(E_{1g})-\omega(E_{2g})=118+300\,\Delta\OA$~\icm, and lies between 107 and 131~\icm{}
for every structure with \OA{} between 1.50 and 1.58~\AA. A pair of measured bands outside
that window is therefore not an E pair. Near equilibrium the $A_{1g}$ band is the most intense,
and remaining ambiguities are removed by the position and strength predicted for the third mode
(Supplementary Table~S1). (iii)~\emph{Inversion.} With three resolved components the
$3\times3$ system is solved exactly for $(\Delta\OA,\Delta\OB,\delta)$. With two components the
asymmetry follows from the axial--basal splitting alone,
$\OB-\OA=24.1~\text{m\AA}+0.41~\text{m\AA}\,[\omega(E_{1g})-\omega(A_{1g})-28.0~\text{cm}^{-1}]$, because the
offset cancels in the difference; the absolute bond lengths then require either the offset
prior or the lattice constants from diffraction. (iv)~\emph{Uncertainty.} The quoted half-width is an
estimated propagated uncertainty under the stated assumptions, not an empirically calibrated
confidence interval. It combines, in quadrature, 0.41~m\AA{} per \icm{} of splitting uncertainty
(an assumed $\sigma=3$~\icm, doubled), the functional spread of the reference asymmetry
($\pm0.8$~m\AA), the closure spread of $J$ ($\pm0.6$~m\AA), the offset prior ($\pm0.15$~m\AA),
the computed confinement and quasi-harmonic bounds ($\le0.1$~m\AA) and an allowance for the
uncomputed differential phonon--phonon shift ($\pm0.8$~m\AA). The assignment and the component
fit are conditions of the inference rather than terms in it. (v)~\emph{Validity.}
The map is harmonic and linearized about the ideal $P6_3/mmc$ crystal; beyond
$|\Delta O|\approx50$~m\AA{} the phonons should be recomputed at the inverted structure, as was
done for every structure in Table~\ref{tab:decisive}. The inversion remains accurate to 2~m\AA{}
under symmetry-breaking uniaxial in-plane strain of up to $\pm2\%$ when the mean of the split
$E_{1g}$ pair is used, and a resolved E-mode splitting signals such strain (Supplementary
Section~S10). The procedure is implemented in \texttt{invert\_raman.py} in the data deposit.

\textbf{Bonding and elasticity.} COHP/ICOHP from LOBSTER~5.1.1~\cite{Nelson2020} (PBE
wavefunctions at r$^2$SCAN+rVV10 geometries); elastic constants by finite strains. Strained-graphite G
band from $3\times3\times2$ graphite supercells at fixed $c$ with in-plane strain $-3\%$ to
$+3\%$.

\textbf{Re-analysis of published data.} Peak positions, HRTEM statistics, Rietveld parameters and
spectra from refs.~\citenum{Yang2025,Lai2026,Yuan2025,Chen2025,Yoshiasa2003,Goryainov2018} including Extended Data,
Supplementary Information and source data. The $\chi^2(z)$ profile re-analysis of the Lai pattern
(Supplementary Information) uses pseudo-Voigt intensity extraction with bootstrap covariance and
a March--Dollase texture proxy~\cite{Dollase1986}. No new experiments were performed.

\section*{Data availability}
The dataset (structures, VASP inputs, phonopy force sets, converged outputs carrying the
energies, stresses, dielectric and elastic tensors, LOBSTER and Bader outputs, refined-structure
jobs, the assignment- and amplitude-validation runs, and all analysis and figure scripts;
2{,}611 files with a self-checking manifest) is
deposited on Zenodo at DOI 10.5281/zenodo.21712712; the calculations added for this revision
(the stress-state map, the strained-cell phonons, the Raman tensors at the refined coordinates,
the quasi-harmonic phonons, the interface cells, and the inversion, force-constant, confinement
and figure scripts; 2{,}228 files with a manifest) are deposited as a new version of the same record (DOI
10.5281/zenodo.22776332). VASP PAW potentials are excluded by licence and identified by name in the Supplementary
Information. Source Data are provided with this paper.

\section*{Code availability}
All analysis scripts are included in the data package.

\section*{Acknowledgements}
Computational resources were provided by the Amarel cluster of the Rutgers Office of Advanced
Research Computing (OARC).

\section*{Funding}
Acknowledgment is made to the Donors of the American Chemical Society Petroleum Research Fund
for partial support of this research (PRF \#70434-ND6).

\section*{Author contributions}
L.Z. designed and performed the research, analysed the data and wrote the manuscript.

\section*{Competing interests}
The authors declare no competing interests.

\bibliographystyle{unsrtnat}
\bibliography{references}

\clearpage
\setcounter{section}{0}\setcounter{table}{0}\setcounter{figure}{0}\setcounter{equation}{0}
\renewcommand{\thesection}{S\arabic{section}}
\renewcommand{\thetable}{S\arabic{table}}
\renewcommand{\thefigure}{S\arabic{figure}}
\begin{center}\LARGE\textbf{Supplementary Information}\end{center}
\vspace{1ex}

\section{Raman inversion: model, identifiability and uncertainty budget}

\textbf{Model.} Linearized mode map about the relaxed structure
($\OA^0=1.5366$, $\OB^0=1.5607$~\AA):
\begin{equation}
\omega_i(\OA,\OB)=\omega_i^0+J_i\cdot(\Delta\OA,\Delta\OB)+\delta ,
\end{equation}
with reference frequencies $\omega^0=(1211.0,\,1301.2,\,1329.2)$~\icm{} for
$(E_{2g},A_{1g},E_{1g})$ and sensitivity matrix (\icm~\AA$^{-1}$)
\begin{equation}
J_{E_{2g}}=(-3025,\,+300),\quad J_{A_{1g}}=(-300,\,-2156),\quad J_{E_{1g}}=(-2725,\,+300),
\end{equation}
from the fixed-bond response surface (Sec.~\ref{sec:surface}). $\delta$ is a common
\emph{additive} frequency offset (nuisance) bounded by the CD/graphite benchmarks computed with
the identical workflow (CD $T_{2g}$ $-6.5$~\icm; graphite G $\approx-13$~\icm{} vs experiment):
prior $|\delta|\le15$~\icm. A multiplicative scale $s$ at the benchmarked $1\%$ level alters a
28~\icm{} splitting by only $\sim$0.3~\icm{} and is subsumed in $\delta$ to leading order.

\textbf{Identifiability.} With two observed components ($y_{A}=1310$, $y_{E}=1338$~\icm, Lai) and
three unknowns $(\Delta\OA,\Delta\OB,\delta)$ the system is underdetermined by one dimension. The
one-parameter solution family is, however, nearly parallel to the $(\Delta\OA,\Delta\OB)$
diagonal: solving the two equations at fixed $\delta$,
\begin{equation}
\frac{\partial(\OB-\OA)}{\partial\delta}=-0.005~\text{m\AA{} per \icm},\qquad
\frac{\partial\OA}{\partial\delta}\approx\frac{\partial\OB}{\partial\delta}
\approx+0.41~\text{m\AA{} per \icm}.
\end{equation}
The \emph{asymmetry} is therefore identified nearly independently of the offset prior (total
prior-induced spread $<0.15$~m\AA{} over $|\delta|\le15$), while the \emph{absolute} bond lengths
inherit $\pm6$~m\AA{} from it. The asymmetry is controlled by the measured splitting:
$\partial(\OB-\OA)/\partial(\text{splitting})=0.41$~m\AA{} per \icm.

\textbf{Measured-cell consistency.} The inversion is performed on the sensitivity map computed
at the calculated lattice constants. Enforcing instead the measured cell of the phase-pure
sample ($a=2.5179$, $c=4.1828$~\AA), which reduces the structural freedom to the internal
coordinate alone, and solving the two component equations for $(z,\delta)$ gives $z=0.0625$,
$\OA=1.545$, $\OB=1.569$~\AA{} and $\OB-\OA=24.0$~m\AA: the identical asymmetry, with both bonds
shifted together by $+12$~m\AA. The required offset grows to $\delta=+28.7$~\icm, outside the
benchmark-derived prior, for an identifiable reason: the measured cell is 1.6\% larger in volume
than the calculated one ($a=2.5024$, $c=4.1679$~\AA), and mode-Gr\"uneisen softening with
$\gamma=1.0$--$1.5$ predicts a uniform $21$--$31$~\icm{} downshift of the computed frequencies
at the measured cell, which the fitted $\delta$ absorbs; the benchmarks that set the
$\pm15$~\icm{} prior were computed at DFT geometries and do not contain this cell term. The
asymmetry is invariant along this family because it is fixed by the splitting (Eq.~3); only the
absolute bond lengths trade off against $\delta$. Exact finite-displacement phonons computed at
this measured-cell structure confirm the linearized treatment:
$E_{2g}/A_{1g}/E_{1g}=1{,}189.0/1{,}280.6/1{,}308.8$~\icm, a near-uniform $-21$~\icm{} shift
from the calculated-cell equilibrium values (inside the Gr\"uneisen estimate), with an
$E_{1g}-A_{1g}$ splitting of 28.2~\icm{} against the measured 28.

\textbf{Assumptions of the inversion.} The procedure rests on five explicit assumptions.
(i)~The scattering volume is dominated by hexagonal diamond in the ideal $P6_3/mmc$ structure, so
that one internal coordinate and two lattice constants describe it. Stacking faults and
interfaces are assumed to be a minority of the probed volume, so that they broaden the bands
rather than shift the fitted component positions. The fault and interface models (Section~S11
and main-text Fig.~2) show that their structural perturbation is confined to a few planes, but
their spectroscopic weight in a given sample is not computed here. (ii)~The harmonic, linearized mode map is accurate over the
inverted range ($|\Delta O|\lesssim50$~m\AA); beyond it the phonons are recomputed exactly, as
for every structure in main-text Table~1. (iii)~The three Raman-active modes retain their
identities. The assignment of the measured components to them is the one preferred by the
E-splitting window and the intensity ordering within the HD spectral models examined (assignment
matrix below), and the inversion is conditional on it. (iv)~Systematic frequency shifts common to all
modes (calibration, functional scale, cell volume, confinement, temperature) are absorbed by the
offset $\delta$; their computed differential parts are given in Section~S13, together with an
allowance for the uncomputed differential phonon--phonon shift. (v)~The average residual stress
state is close to hydrostatic: the published lattice constants indicate a few tenths of a per
cent of strain, and no E-mode splitting is resolved (main text; Section~S10). Local strains are
not bounded by these averages, but Section~S10 shows that the inversion tracks the true asymmetry
to 2~m\AA{} over $\pm2\%$ of uniaxial in-plane strain. Under these
assumptions the inverse problem is low-dimensional: three structural degrees of freedom, of
which diffraction fixes two, against three mode positions.

\textbf{Uncertainty budget for $\OB-\OA$.} The quoted half-width is an estimated propagated
uncertainty under the stated assumptions, not an empirically calibrated confidence interval. Each
entry is a two-standard-deviation-equivalent half-width under its assumption. The entries are
treated as independent and combined in quadrature, and the model choices (assignment, line shape)
are conditions of the inference rather than terms in it. The entries are: component-fit
uncertainty of the
splitting (assumed $\sigma\approx3$~\icm, doubled) $\rightarrow$ $\pm2.4$~m\AA; functional
dependence of the reference asymmetry (24.1 r$^2$SCAN+rVV10 vs 23.4 PBE) $\rightarrow$ $\pm0.8$;
closure spread of $J$ (7\%) acting on the inferred correction $\rightarrow$ $\pm0.6$; offset
prior $\rightarrow$ $\pm0.15$; differential phonon-confinement shift (computed) $\rightarrow$
$\le\pm0.06$; differential 300-K quasi-harmonic shift (computed) $\rightarrow$ $\le\pm0.01$;
allowance for the uncomputed differential phonon--phonon shift, taken as large as the whole
300-K shift of the diamond line (2~\icm) $\rightarrow$ $\pm0.8$ (Section~S13). Combined:
\begin{equation}
\boxed{\;\OB-\OA = 24\pm3~\text{m\AA{} (estimated propagated uncertainty), phase-pure sample}\;}
\end{equation}
excluding zero, Yang's $-55$~m\AA{} and Lai's $+183$~m\AA{} by wide margins (zero lies eight
half-widths from the central value). Lai \textit{et al.}\ do not publish
component-fit uncertainties; $\sigma\approx3$~\icm{} is our estimate for the splitting of two
narrow, separately fitted components whose positions are quoted to 1~\icm. The result scales
linearly with this choice: doubling it to $\sigma=6$~\icm{} widens the half-width to $\pm5$~m\AA{}
and changes no conclusion. The interval is conditional on the $A_{1g}$/$E_{1g}$
assignment, which is fixed independently (assignment matrix below).

\textbf{Scope of this interval.} The budget above propagates the uncertainties of \emph{one}
measurement (the two narrow, separately fitted components of the phase-pure sample), and the
interval should be quoted as such. It is not a between-sample interval. The independent lonsdaleite
spectrum of Goryainov \textit{et al.} provides the only available external test: their
$A_{1g}=1{,}305$~\icm{} lies 4~\icm{} from the relaxed prediction, but their
$A_{1g}$--$E_{1g}$ splitting of 51~\icm{} (against 28 predicted) maps through
$\partial(\OB-\OA)/\partial(\text{splitting})=0.41$~m\AA{} per \icm{} to $\OB-\OA\approx33$~m\AA.
Those authors report band widths of 50--60~\icm{} and a $\sim$10~\icm{} low-energy shift from
lonsdaleite imperfection, on a spectrum obtained by subtracting two spectra of a mixture
containing 0--42\% lonsdaleite. Taking the component-position uncertainty as a quarter to a third
of the band width gives a splitting uncertain to 18--28~\icm{} (we adopt 20), hence
$\OB-\OA=33\pm8$~m\AA, an interval that overlaps the phase-pure one. The three independent
experimental determinations therefore read $+24\pm3$, $+33\pm8$ and $+60\pm45$~m\AA: overlapping,
all positive, and none compatible with either recent refinement. Their uncertainties are not on
a common statistical footing (the first two are propagated estimates of this section, the third
is the refinement's own bond error), so no formal significance is attached to their agreement.

\textbf{Assignment matrix.} All six ordered assignments of the two measured components to the
three modes were inverted:

\begin{table}[h]\centering\footnotesize
\setlength{\tabcolsep}{4pt}
\caption{Assignment matrix for Lai's two measured components (1{,}310/1{,}338~\icm).}
\begin{tabular}{lccccL{3.4cm}}
\toprule
assignment & $\OA$ (\AA) & $\OB$ (\AA) & $\OB-\OA$ (m\AA) & implied state & third-mode prediction \\
\midrule
$A_{1g}$/$E_{1g}$ & 1.533 & 1.557 & $+24$ & low stress & $E_{2g}$ 1{,}221 (in envelope tail) \\
$A_{1g}$/$E_{2g}$ & 1.495 & 1.563 & $+67$ & $\sim$42~GPa in-plane & $E_{1g}$ 1{,}443 (unobserved) \\
$E_{2g}$/$A_{1g}$ & 1.503 & 1.549 & $+46$ & $\sim$30~GPa in-plane & $E_{1g}$ 1{,}418 (unobserved) \\
$E_{1g}$/$A_{1g}$ & 1.542 & 1.543 & $+1.2$ & near-equilibrium, low stress & $E_{2g}$ 1{,}191 (exact; unobserved) \\
$E_{2g}$/$E_{1g}$ & 1.237 & -- & -- & unphysical ($\Delta\OA=-300$~m\AA) & -- \\
$E_{1g}$/$E_{2g}$ & 1.049 & -- & -- & unphysical ($\Delta\OA=-487$~m\AA) & -- \\
\bottomrule
\end{tabular}
\end{table}

Two assignments are unphysical outright. Two ($A_{1g}$/$E_{2g}$ and $E_{2g}$/$A_{1g}$) require
tens-of-GPa in-plane stress states excluded by the ambient lattice constants and predict a
strong $E_{1g}$ band where nothing is observed. The remaining alternative, $E_{1g}$/$A_{1g}$
(the mode ordering inverted), is mechanically innocuous: it inverts to a near-symmetric
structure ($\OB-\OA=1.2$~m\AA; 1.16--1.31 over $|\delta|\le15$~\icm) close to equilibrium, and
neither energy nor stress excludes it. It is excluded by the spectra, on three counts.
(i)~Its third mode falls at 1{,}191~\icm{} ($E_{2g}$), in-window and unobserved. This is the exact
finite-displacement value at the swap structure, which also reproduces the two components at
1{,}312/1{,}338~\icm{} and so confirms the linearized inversion to $\le2$~\icm. Exact finite-field
Raman tensors at the same structure (method of Sec.~\ref{sec:intensity}) give degeneracy-summed
activities $E_{2g}{:}A_{1g}{:}E_{1g}=98{:}100{:}103$. At this near-equal-bond geometry the three
bands are nearly equal in activity, so the unobserved 1{,}191~\icm{} band is predicted to be as
strong as the two observed components.
(The single-variable activity interpolation of Sec.~\ref{sec:intensity}, valid along the scan,
fails in this bond-crossover region; the exact tensors supersede it here.)
(ii)~The same near-equal activities give no account of the strongly asymmetric measured
envelope: its dominant 1{,}310~\icm{} component is reproduced by the preferred assignment as
the dominant $A_{1g}$ ($100{:}41{:}28$ at the relaxed structure), while the inverted reading
predicts two components of equal strength. (iii)~On the one sample
where all three modes are resolved (Goryainov \textit{et al.}), the E-splitting bound of
Sec.~\ref{sec:lsq} admits only the pairing $E_{2g}=1{,}244$/$E_{1g}=1{,}356$ (splitting
112~\icm, inside 107--131; the alternative pairings give 61 and 51), which forces
$A_{1g}=1{,}305$ as the middle, most intense band, i.e.\ the normal ordering; the same ordering
appears in the simulated spectrum of Lai \textit{et al.} Under the inverted reading the
three-determination concordance of this section would also collapse (a $+1$~m\AA{} value against
$+33\pm8$). The $A_{1g}$/$E_{1g}$ assignment is thus preferred, within the HD spectral models examined, by
intensity ordering, the third-mode constraint and the resolved third-sample spectrum. It implies a
scattering geometry (or texture distribution) with $E_{1g}$ visibility; the rejections that
invoke an unobserved $E_{1g}$ band apply under that same visibility condition, so the analysis is
self-consistent. The quantitative inversion is conditional on this assignment and on that
visibility. The activities entering these rejections are powder-averaged, static and
non-resonant, whereas the measurement was made on oriented material under ultraviolet
excitation.

\textbf{Scattering geometries.} $D_{6h}$ Raman tensors (calculated, arbitrary units):
$A_{1g}$: $\alpha_{xx}=\alpha_{yy}=-1.31$, $\alpha_{zz}=+6.53$;
$E_{2g}$: $(\alpha_{xx}-\alpha_{yy},\alpha_{xy})$ components $2.15$;
$E_{1g}$: $(\alpha_{xz},\alpha_{yz})$ components $1.78/0.48$.
Visibility by geometry: backscattering along $c$: $A_{1g}$ (via $\alpha_{xx}$), $E_{2g}$ allowed;
$E_{1g}$ forbidden. Edge-on with in-plane analyzer: $A_{1g}$, $E_{2g}$. Edge-on with a $c$-axis
polarization component: all three, $A_{1g}$ strongly enhanced via $\alpha_{zz}$. Powder average:
all three with degeneracy-summed activities $100{:}41{:}28$ ($A_{1g}{:}E_{2g}{:}E_{1g}$) at the
relaxed geometry. Activities are static, non-resonant (PBE); both experiments used UV excitation
where resonance may reweight intensities; frequencies are the primary evidence throughout.

\section{Fixed-bond response surface and constructions}\label{sec:surface}

Structures with $(\OA,\OB)$ fixed were generated by the exact geometric closure
$z=(c/2-\OB)/(2c)$, $a=\sqrt{3(\OA^2-(c/2-\OB)^2)}$, with the remaining degree of freedom $c$
resolved either by energy minimization over a 5-point scan (E-min closure) or fixed to the
experimental value (fixed-$c$ closure). Sensitivities were obtained from the fixed-\OA{} path
($\OA=1.58$, $\OB=1.44$--$1.60$~\AA), the fixed-\OB{} pair ($\OB=1.5607$, $\OA=1.50/1.58$), and
the fixed-cell $z$ path; the $A_{1g}$ \OB-sensitivity is $-2000\pm150$~\icm~\AA$^{-1}$ across all
three constructions (the constructions differ in which cell parameters relax; the spread bounds
the closure dependence).

\section{Physical state of all non-equilibrium structures}

\begin{table}[h]\centering\footnotesize
\setlength{\tabcolsep}{3.5pt}
\caption{Physical-state data for every constrained structure. Stresses are positive-compressive
(the confining stress required to hold the structure); residual forces are per-atom maxima in the
undisplaced supercell before subtraction. No structure has imaginary $\Gamma$ modes. Stress sign: positive = compressive (confining); negative = tensile.}
\begin{tabular}{lcccccccc}
\toprule
structure & $a$ (\AA) & $c$ (\AA) & $z$ & $\OA/\OB$ (\AA) & $\Delta E$ (meV/at) & $\sigma_{xx}$ & $\sigma_{zz}$ & $F_{\max}$ (eV/\AA) \\
\midrule
relaxed & 2.5024 & 4.1679 & 0.0628 & 1.537/1.561 & 0 & 0 & 0 & 0.00 \\
$z$-scan $z{=}0.063$ & 2.5024 & 4.1679 & 0.0630 & 1.537/1.559 & $\sim$0 & -- & -- & 0.07 \\
Yang Rietveld & 2.5182 & 4.1780 & 0.0693 & 1.565/1.510 & 31 & $-8.5$ & $-2.0$ & 2.18 \\
Lai Rietveld & 2.5179 & 4.1828 & 0.0479 & 1.508/1.691 & 122 & $-5.8$ & $+8.2$ & 3.54 \\
Lai mixed & 2.5150 & 4.1720 & 0.0504 & 1.512/1.666 & 87 & $-4.5$ & $+7.5$ & 3.13 \\
$z$-scan $\OB{=}1.442$ & 2.5024 & 4.1679 & 0.0770 & 1.581/1.442 & 149 & $+0.3$ & $+10.3$ & 5.30 \\
fixed-$\OA{=}1.58$, E-min & 2.5423 & 4.0494 & 0.0722 & 1.580/1.440 & 104 & $-20.6$ & $+54.1$ & 3.76 \\
fixed-$\OA{=}1.58$, exp-$c$ & 2.4951 & 4.1780 & 0.0777 & 1.580/1.440 & 164 & $+4.2$ & $+6.9$ & 5.55 \\
fixed-$\OA{=}1.56$, E-min & 2.5136 & 4.0245 & 0.0711 & 1.560/1.440 & 90 & $-7.0$ & $+64.7$ & 3.42 \\
fixed-$\OA{=}1.56$, exp-$c$ & 2.4571 & 4.1780 & 0.0777 & 1.560/1.440 & 179 & $+24.9$ & $+6.1$ & 5.61 \\
compressed LSQ & 2.4640 & 3.9904 & 0.0674 & 1.521/1.457 & 81 & 22.1 & 77.3 & 1.95 \\
\bottomrule
\end{tabular}
\end{table}

$\Delta E$ values for the Yang/Lai refined structures are static energies at the published
coordinates relative to relaxed. Residual-force subtraction (phonopy \texttt{--fz}) was validated
at the equilibrium structure ($<$4~\icm{} against the unconstrained calculation); frequencies of
strongly non-stationary structures are curvatures of constrained structural models, not
predictions for free-standing phases, which is precisely their role in this analysis.

\section{Two-bond least-squares search: construction, residuals and\\ uniqueness}\label{sec:lsq}

\textbf{Construction.} The search asks whether \emph{any} homogeneous $P6_3/mmc$ structure
reproduces all three bands reported by Yang \textit{et al.} (1{,}249, 1{,}321, 1{,}529~\icm). Over
the $(\OA,\OB)$ manifold we evaluate the linearized mode map of Sec.~1 (Eqs.~1--2) on a grid, and
score each point by the offset-minimized root-mean-square misfit
\begin{equation}
R(\OA,\OB)=\min_{|\delta|\le15}\sqrt{\tfrac{1}{3}\textstyle\sum_i\big[\omega_i(\OA,\OB)+\delta-y_i\big]^2},
\end{equation}
with $y=(1249,1321,1529)$~\icm{} assigned to $(E_{2g},E_{1g},A_{1g})$ and $\delta$ the same common
additive nuisance offset used in Sec.~1, bounded by the CD/graphite benchmarks. The resulting
misfit surface is Fig.~5a of the main text. It has a single connected low-misfit island, doubly
compressed relative to equilibrium; a point inside it was then recomputed with exact
finite-displacement DFT rather than the linearization.

\textbf{Exact check inside the low-misfit island.} The bounded-offset minimum of Eq.~(5) on the
grid is at $(\OA,\OB)=(1.514,1.451)$~\AA{} ($R=16.2$~\icm). Exact phonons were computed at a
point inside the island, $(\OA,\OB)=(1.521,1.457)$~\AA{} (linearized $R=17.0$~\icm, within
1~\icm{} of the minimum), with the energy-minimized cell ($a=2.4640$, $c=3.9904$~\AA,
$z=0.06744$); the exact $\Gamma$ phonons there are
$E_{2g}=1{,}233$, $E_{1g}=1{,}345$, $A_{1g}=1{,}552$~\icm, against the linearized prediction
(1{,}230/1{,}343/1{,}529). Per-band residuals against the measured values are $+16$, $-24$ and
$-23$~\icm{} at $\delta=0$, i.e.\ every band matched to within the 15--30~\icm{} scale error. The
offset-minimized RMS misfit is 18.6~\icm{} at $\delta=-10.3$~\icm.

\textbf{Residual structure.} The residual is not random but dominated almost entirely by the E-mode
splitting. From the sensitivity matrix of Sec.~1, the splitting is controlled by \OA{} alone
to linear order, the \OB{} cross-terms cancelling exactly:
\begin{equation}
\omega(E_{1g})-\omega(E_{2g}) = 118.2 + 300\,\Delta\OA~\text{\icm{} (}\Delta\OA\text{ in \AA)} .
\end{equation}
The least-squares structure therefore predicts an E-splitting of 112~\icm{} where 72~\icm{} is
observed; the $+16$ and $-24$~\icm{} per-band residuals are that single 40~\icm{} discrepancy
distributed over two bands by the fit. This is worth stating plainly because it bounds how good any match of this
kind can be.

\textbf{Uniqueness.} Eq.~(6) also settles which assignments are admissible, without any further
calculation. Across the entire range of \OA{} sampled in this work (1.50--1.58~\AA) the predicted
E-mode splitting is confined to 107--131~\icm. The three ways of assigning two of the measured
bands to the two E modes give observed splittings of 72~\icm{} (1{,}249/1{,}321), 208~\icm{}
(1{,}321/1{,}529) and 280~\icm{} (1{,}249/1{,}529). Only the first lies within 40~\icm{} of the
accessible band; the other two are excluded by 77 and 149~\icm{} respectively, several times the
scale error, for \emph{every} structure on the manifold. The assignment used above is thus the only
one worth searching under, and the low-misfit island it selects is unique within it. The
complementary four-way assignment analysis for the two-component spectrum of Lai \textit{et al.}\
is given in Sec.~1.

\textbf{Disposition.} The existence of this solution refutes the narrower claim that no HD
structure reproduces all three bands, and the main text states the stronger claim instead: the
matching structure lies 81~meV\,atom$^{-1}$ above equilibrium, requires $\sigma_{xx}=22.1$ and
$\sigma_{zz}=77.3$~GPa of confining stress (Table~S2), and has lattice constants 2.2\% and 4.5\%
smaller than the ambient values measured on the same sample. It describes hexagonal diamond under
tens of gigapascals, not a recovered ambient specimen.

\textbf{Surrogate documentation.} The misfit surface of main-text Fig.~5a evaluates Eq.~(5)
exactly as written, with the offset clipped to $|\delta|\le15$~\icm{} at every grid point. The
clip matters: without it a spurious low-misfit diagonal band appears, along which offsets of
hundreds of \icm{} would compensate the structure; with it, only 3\% of the plotted manifold lies
below 30~\icm. The accuracy of the linearization at the far point is bounded by the exact check
above (largest deviation 23~\icm, for $A_{1g}$, where the linearization under-predicts the
frequency). The energy surface of Fig.~5b is $E_{\min}(\OA,\OB)$, the energy minimized over the
remaining cell parameter at fixed bonds. It is represented by the quadratic form
$E=k_{AA}\,\Delta\OA^2+2k_{AB}\,\Delta\OA\,\Delta\OB+k_{BB}\,\Delta\OB^2$, least-squares fitted to
the six computed structures with that closure: relaxed; fixed-$\OA$ 1.58 and 1.56~\AA{} at
$\OB=1.44$~\AA, E-min; fixed-$\OB$ with $\OA=1.50$ and 1.58~\AA; and the exact-check point of
Table~S2. The fit gives $(k_{AA},k_{AB},k_{BB})=(21{,}300,\,2{,}700,\,6{,}400)$~meV\,atom$^{-1}$\,\AA$^{-2}$,
with rms residual 2~meV\,atom$^{-1}$ and maximum residual 4~meV\,atom$^{-1}$ (fit and points in
the archived plotting script). Structures with other closures are not on this surface and lie
above it by construction, because the same $(\OA,\OB)$ can be closed by different cells with
different energies. The fixed-cell internal-coordinate scan lies above it by up to
33~meV\,atom$^{-1}$ at $z=0.075$, and the fixed-$c$ variants of Table~S2 by
60--90~meV\,atom$^{-1}$. The $\gtrsim$80~meV\,atom$^{-1}$
contrast used in the argument refers to $E_{\min}$ and is an order of magnitude above the fit
residual.

\section{Intensity-weighted comparison with the measured spectra}\label{sec:intensity}

Figure~4 of the main text plots predicted mode positions as equal-height sticks, so that the
frequency comparison is not conditioned on calculated intensities. This section supplies the
intensity-weighted reading. Degeneracy-summed band activities were computed at three geometries
spanning the scan (Methods; PBE, static and non-resonant):

\begin{table}[h]\centering\footnotesize
\setlength{\tabcolsep}{5pt}
\caption{Degeneracy-summed relative band activities, normalized to $A_{1g}=100$, from exact
finite-field Raman tensors at each geometry (static, non-resonant, PBE; eigenvectors of each
structure's own residual-force-subtracted force constants). Rows above the rule: geometries of
the internal-coordinate scan; below the rule: refined and candidate structures.}
\begin{tabular}{lcccccc}
\toprule
geometry & \OB{} (\AA) & $E_{2g}$ (\icm) & $A_{1g}$ (\icm) & $E_{1g}$ (\icm) & $E_{2g}$ : $A_{1g}$ : $E_{1g}$ & $\varepsilon_\infty$ \\
\midrule
$z$-scan $z{=}0.055$ & 1.626 & 1{,}299 & 1{,}144 & 1{,}399 & 3 : 100 : 0.4 & 5.99 \\
relaxed & 1.561 & 1{,}211 & 1{,}301 & 1{,}329 & 41 : 100 : 28 & 5.75 \\
$z$-scan $z{=}0.073$ & 1.475 & 1{,}077 & 1{,}472 & 1{,}214 & 27 : 100 : 107 & 5.86 \\
\midrule
Rietveld, Yang \textit{et al.} & 1.510 & 1{,}104 & 1{,}399 & 1{,}238 & 129 : 100 : 317 & 5.79 \\
Rietveld, Lai \textit{et al.} & 1.691 & 1{,}343 & 959 & 1{,}425 & 0.1 : 100 : 1.2 & 6.79 \\
Lai mixed HD/CD model & 1.666 & 1{,}325 & 1{,}032 & 1{,}414 & 0.6 : 100 : 1.5 & 6.36 \\
$z$-scan $\OB{=}1.442$ & 1.442 & 1{,}017 & 1{,}531 & 1{,}161 & 8 : 100 : 51 & 6.03 \\
compressed LSQ point & 1.457 & 1{,}233 & 1{,}552 & 1{,}345 & 120 : 100 : 323 & 5.65 \\
\bottomrule
\end{tabular}
\end{table}

The $A_{1g}$ polarizability derivative $\alpha_{zz}$ changes sign between the relaxed and the
short-bond geometries, so the $A_{1g}$ activity passes through a minimum between them. The
activity ordering therefore cannot be interpolated along the scan and must be computed at each
structure. The readings below use the exact tensors.

\textbf{Relaxed structure} ($\OB=1.561$~\AA). A dominant $A_{1g}$ at 1{,}301~\icm{} overlapped by
$E_{1g}$ at 1{,}329~\icm{} (28\%), with a weaker $E_{2g}$ at 1{,}211~\icm{} (41\%). The predicted
spectrum is one asymmetric envelope with its maximum near 1{,}305~\icm{} and a low-frequency
shoulder, the observed form in both samples and the reason the two fitted components of
Lai \textit{et al.}\ are reproduced to 9~\icm{} each.

\textbf{Yang Rietveld} ($\OB=1.510$~\AA). The $A_{1g}$ mode sits near its activity minimum, and
the strongest predicted band is the $E_{1g}$ at 1{,}238~\icm{} (317\% of $A_{1g}$), followed by
$E_{2g}$ at 1{,}104~\icm{} (129\%) and $A_{1g}$ at 1{,}399~\icm. The near-coincidence of the
1{,}238~\icm{} band with the observed 1{,}249~\icm{} band is thus a strong-band coincidence, but
the pattern is not consistent with the measurement. The strongest measured band (1{,}321~\icm)
has no counterpart, the second-strongest predicted band (1{,}104~\icm) is absent, and
1{,}529~\icm{} is unexplained.

\textbf{Lai Rietveld and mixed models} ($\OB=1.691$ and $1.666$~\AA). Most of the scattering
strength lies in the $A_{1g}$ band at 959 or 1{,}032~\icm, below the published measurement
window; the in-window $E_{2g}$ and $E_{1g}$ bands carry 0.1--0.6\% and 1.2--1.5\% of it. Relative
activities set no absolute detection limit, so this weakness is supporting evidence only. The
decisive comparison is one of positions: the two in-window modes are 82~\icm{} (HD-only) or
89~\icm{} (mixed) apart, whereas the two measured components are 28~\icm{} apart, and no common
offset reconciles the patterns.

\textbf{Short-\OB{} structures} ($\OB=1.442$--$1.457$~\AA). At $\OB=1.442$~\AA{} the $E_{1g}$
band at 1{,}161~\icm{} carries half of the $A_{1g}$ activity, so a locally compressed region would
present 1{,}529~\icm{} together with a band near 1{,}161~\icm{} at half its strength. At the
exact-check point of the compressed island, a homogeneous candidate for all three bands, the
$E_{1g}$ (1{,}345~\icm) and $E_{2g}$ (1{,}233~\icm) bands are predicted stronger than the $A_{1g}$
(1{,}552~\icm), whereas the measured 1{,}529~\icm{} band is weaker than 1{,}321 but stronger than
1{,}249~\icm. The companion band near 1{,}161~\icm{} is the basis of the main-text argument
against a locally compressed HD origin for the 1{,}529~\icm{} feature. For a minority component
only its ratio to the 1{,}529~\icm{} feature itself is diagnostic, not its rank among all bands
of the sample.

Two caveats apply throughout. Activities are static and non-resonant, whereas both experiments used
ultraviolet excitation, where resonance can reweight bands; and they are computed at the PBE level.
Frequencies therefore remain the primary evidence and these orderings are supporting. The
$A_{1g}$-dominant pattern near equilibrium is independently corroborated by the simulated spectrum
published by Lai \textit{et al.}, computed with the same functional class.

\section{Diffraction sensitivity analysis of the published Lai pattern}

\textbf{Data provenance.} Observed, calculated, background and difference curves for Extended
Data Fig.~2a of Lai \textit{et al.} were taken from the publisher-hosted source-data spreadsheet
(623 rows, $2\theta=14.96$--$40.02^\circ$, $\lambda=0.6888$~\AA); no digitization from images was
used. A uniform $+0.08^\circ$ offset of the source data relative to the nominal peak positions
was detected by comparison against the authors' own calculated curve and absorbed into peak
positions.

\textbf{Method.} Ten reflections in range were modelled: (100), (002), (101), (102), (110),
(103), (200), (112), (201), (004). The $hk0$ intensities are $z$-independent (verified analytically) and
anchor scale and $B_{\rm iso}$. Background-subtracted intensities were extracted by pseudo-Voigt
fits (shared width scaling within the overlapping low-angle triplet); uncertainties and the full
$10\times10$ covariance by block bootstrap over the residual noise (300 replicates). The
intensity model used carbon Cromer--Mann form factors [S1], multiplicity, Lorentz-polarization,
$B_{\rm iso}$, and a one-parameter March--Dollase texture proxy along $[001]$. At each $z$ on a
grid $0.040$--$0.080$, (scale, $B_{\rm iso}$, $r_{\rm MD}$) were re-optimized; $\chi^2_{\rm eff}$
is a \emph{sensitivity metric} (generalized least squares against the bootstrap covariance,
inflated by the goodness-of-fit), not a formal likelihood: the original refinement used an
8th-order spherical-harmonic texture model (texture index 5.5) that a one-parameter proxy cannot
reproduce, and no conventional $\Delta\chi^2$ confidence interpretation is claimed.

\begin{figure}[h]\centering
\includegraphics[width=120mm]{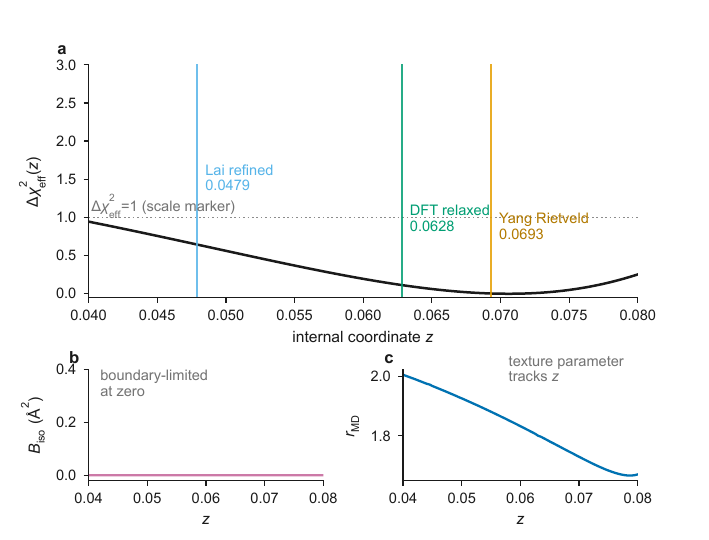}
\caption{Sensitivity profile $\Delta\chi^2_{\rm eff}(z)$ for the published Lai pattern with the
three coordinates marked, and the accompanying $B_{\rm iso}(z)$ and $r_{\rm MD}(z)$ traces
showing the parameter degeneracy.}
\label{fig:zrefit}
\end{figure}

\begin{figure}[h]\centering
\includegraphics[width=\textwidth]{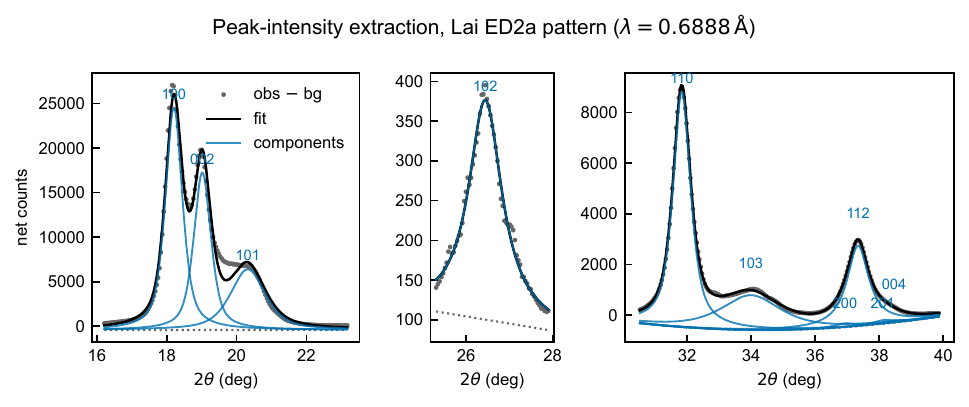}
\caption{Quality control on the intensity extraction. Windowed pseudo-Voigt fits to the
background-subtracted source data of Extended Data Fig.~2a of Lai \textit{et al.}, with shared
width scaling inside the overlapping low-angle triplet and free widths for the fault-broadened
(101) and (103) reflections. The extracted integrated intensities and their block-bootstrap
covariance are the input to the $\chi^2(z)$ profile of Fig.~\ref{fig:zrefit}.}
\label{fig:zrefitqc}
\end{figure}

\textbf{Result.} The profile is flat: all $z\in[0.040,0.080]$ lie within
$\Delta\chi^2_{\rm eff}\le1$ of the minimum ($z=0.0705$; minimum position shifts to 0.0645 under
a diagonal-covariance treatment, itself a degeneracy signature). $B_{\rm iso}$ pins at the
zero bound across the grid and $r_{\rm MD}$ tracks $z$, the classic ill-conditioning signatures;
the most $z$-sensitive reflections ($00l$) are exactly those most affected by the strong texture.
In every treatment examined, the fit at $z\approx0.063$--$0.070$ is at least as good as at the
refined $z=0.0479$. Representative fits at $z=0.0479$, $0.0628$ and the minimum, the extracted
intensity table, covariances and parameter traces are archived with the analysis script
(\texttt{zrefit\_lai.py}). The sensitivity profile itself is shown in Fig.~\ref{fig:zrefit}, and
the peak-extraction quality-control overlay in Fig.~\ref{fig:zrefitqc}.

\FloatBarrier
\section{Strained-sp$^2$ candidate for the high-frequency band}

\begin{figure}[h]\centering
\includegraphics[width=120mm]{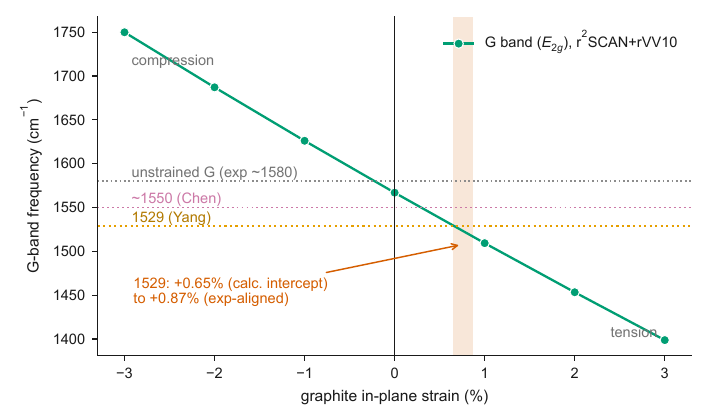}
\caption{Graphite G band versus in-plane strain ($-58.5$~\icm~\%$^{-1}$; unstrained calculated
1{,}567 vs experimental $\approx$1{,}580~\icm). The 1{,}529~\icm{} band of Yang \textit{et al.}
corresponds to $+0.65$--$0.87\%$ tension depending on intercept calibration; the
$\approx$1{,}550~\icm{} band of Chen \textit{et al.} to $+0.3$--$0.5\%$. This establishes
plausibility of a strained minority-sp$^2$ origin; it does not identify the carrier (no interface
model, resonance profile or linewidth calculation is implied).}
\label{fig:gband}
\end{figure}

\FloatBarrier
\section{Numerical convergence and validation}

Computed with the identical workflow: CD $T_{2g}$ $=1{,}325.5$ vs measured $1{,}332$~\icm{}
($-0.5\%$; $4\times4\times4$ supercell) and graphite G $=1{,}567$ vs $\approx$$1{,}580$~\icm{}
($-0.8\%$; $3\times3\times2$) bound the absolute frequency scale near equilibrium at $\sim$1\%,
consistent with the PBE-vs-r$^2$SCAN+rVV10 spread ($12$--$26$~\icm) on identical structures.
Residual-force subtraction reproduces the unconstrained equilibrium frequencies to $<4$~\icm.
The $z$-scan and refined-structure phonons use the same $3\times3\times2$ supercell,
$7\times7\times5$ k-mesh, 800~eV cutoff and $10^{-8}$~eV SCF threshold as the validated
equilibrium calculation; phonopy displacement amplitude 0.01~\AA{} (default), Raman-tensor
finite-field displacements $\pm0.02$~\AA. Symmetry assignments track irreducible representations
computed at each geometry (no nearest-frequency relabeling across the scan). At strongly
non-stationary structures (residual forces up to 5.6~eV\,\AA$^{-1}$, Table~S2) the leading
finite-displacement error of a \emph{one-sided} difference grows with the residual force through
cubic anharmonicity; the calculations here are protected against it by construction, since all
force constants are built from symmetric plus--minus displacement pairs (verified in the
archived displacement metadata of the equilibrium, scan and refined-structure runs), which
cancel the cubic term and leave a quartic-order residual. A direct test at the two
highest-residual-force structures (2.2 and 5.6~eV\,\AA$^{-1}$: the Yang-Rietveld and
fixed-$\OA{=}1.56$/experimental-$c$ rows of Table~S2) confirms this: halving the displacement
amplitude to 0.005~\AA{} with forced plus--minus pairs reproduces every Raman-active frequency
to 0.1~\icm. The discriminating margins in the main text (78--350~\icm) in any case exceed the
equilibrium-validated scale error by an order of magnitude. All inputs are archived, so the $2\times2\times2$ supercell cross-check and the
displacement-amplitude sensitivity test can be reproduced directly from the deposited data.

\textbf{One-phonon spectral bound.} The full phonon density of states of relaxed HD, computed
from the archived equilibrium force constants on a $24\times24\times16$ wavevector mesh, has
support up to 1{,}329~\icm{} only; the top of the one-phonon spectrum is the zone-centre
$E_{1g}$ mode. No one-phonon state of the relaxed structure, at any wavevector, lies within
200~\icm{} of the 1{,}529~\icm{} band discussed in the main text. The density of states is
regenerable from the deposited \texttt{FORCE\_SETS} with a single phonopy mesh calculation.

\section{Scope of spectral comparisons}

All comparisons in the main text are made at the level of \emph{reported band and fitted
component positions}; no digitized experimental traces were used. Statements about envelope
asymmetry and unresolved components refer to the published figures qualitatively and are worded
accordingly. The underlying Raman traces of Lai \textit{et al.}\ (their Supplementary Fig.~4)
are not among that article's deposited source-data files, so component-fit line shapes and
covariances are not independently accessible. The $\sigma\approx3$~\icm{} adopted in Sec.~S1,
and its factor-of-two sensitivity bound, parametrize that dependence explicitly.

{\sloppy The archived data package (2{,}611 files, with a full manifest; Zenodo DOI
10.5281/zenodo.21712712) contains the structures
(\texttt{POSCAR} and \texttt{CONTCAR} files) and VASP inputs for every calculation reported here; the
phonopy force sets and displacement metadata; the converged \texttt{OUTCAR}s carrying the total
energies, stress tensors, dielectric tensors and elastic constants quoted in the text and in
Table~S2; the LOBSTER COHP/ICOHP outputs behind the bonding analysis; the Bader and ELF outputs;
the validation runs of this Supplementary Information (the assignment-check, measured-cell,
displacement-amplitude and swap-structure Raman-tensor calculations);
and every analysis and plotting script, including those that generate the published figures. The
calculations added for the revision (Sections~S10--S13 and main-text Fig.~2: the 38 stress
states, the strained-cell phonons, the Raman tensors at the refined coordinates, the
quasi-harmonic phonons, the fourteen interface cells, and the inversion, force-constant,
confinement and figure scripts; 2{,}228 files with a manifest) are deposited as a new version of the same record (DOI
10.5281/zenodo.22776332). Two classes of file are deliberately excluded: the VASP PAW potentials (\texttt{PAW\_PBE C 08Apr2002},
VASP 6.4.2), whose licence forbids redistribution and which are identified by name instead; and
charge densities and wavefunctions, which are regenerable from the included inputs. Everything
required to reproduce the numbers and figures of this work is present, subject to supplying the
licensed potential.\par}

\FloatBarrier
\section{Stress-state map in the observed orientation relationship}\label{sec:strainmap}

The states of main-text Fig.~2 were generated by applying Cartesian strain tensors to the relaxed
cell and relaxing the internal coordinates at fixed cell. ``Clamped'' states hold the transverse cell dimensions. ``Free'' states set the transverse
strains from the computed elastic tensor ($C_{11}=1{,}255$, $C_{12}=98$, $C_{13}=8$,
$C_{33}=1{,}366$~GPa), which left every transverse stress below 0.3~GPa, so no correction round
was needed. Table~\ref{tab:uniY} lists the uniaxial strain along $[10\bar{1}0]$, the image of the
graphite $[001]$ formation-stress axis under the orientation relationship
G$(0001)\parallel$HD$(10\bar{1}0)$, G$[10\bar{1}0]\parallel$HD$[0001]$. The strain lowers the
symmetry to orthorhombic and separates the basal bond whose projection lies along the strain axis
(the bond spanning the former graphite gallery) from the two in-sheet basal bonds. Compression along this axis enlarges the asymmetry; only tension above $\approx+2\%$ lets
the single gallery-spanning bond overtake \OB, while the mean asymmetry stays positive to $+3\%$.
Uniaxial strain along $[2\bar{1}\bar{1}0]$ gives the same mean asymmetry to within 1~m\AA{}
($+36$, $+30$, $+18$, $+11$~m\AA{} at $-2$, $-1$, $+1$, $+2\%$). The transverse-free inversion
thresholds for uniaxial-$c$ and biaxial-$a$ strain are $-1.77$ and $+1.82\%$ (clamped $-1.79$
and $+1.84\%$). The complete 38-state table is in the Source Data.

\begin{table}[h]\centering\footnotesize
\setlength{\tabcolsep}{5pt}
\caption{Uniaxial strain along HD$[10\bar{1}0]$ (transverse dimensions clamped; the
transverse-free mean asymmetry is given in the last column). $O_A^{\rm g}$ is the gallery-spanning
basal bond, $O_A^{\rm s}$ the mean of the two in-sheet basal bonds; $\sigma_\parallel$ is the
stress along the strain axis (compressive positive).}
\label{tab:uniY}
\begin{tabular}{rcccccccc}
\toprule
$\varepsilon$ (\%) & $\sigma_\parallel$ (GPa) & \OB{} (\AA) & $O_A^{\rm g}$ (\AA) & $O_A^{\rm s}$ (\AA) & $\langle\OA\rangle$ (\AA) & $\OB-\langle\OA\rangle$ (m\AA) & $\OB-O_A^{\rm g}$ (m\AA) & free (m\AA) \\
\midrule
$-3$ & 40.8 & 1.5591 & 1.5008 & 1.5250 & 1.5169 & $+42.2$ & $+58.3$ & $+41.0$ \\
$-2$ & 26.4 & 1.5598 & 1.5123 & 1.5289 & 1.5234 & $+36.4$ & $+47.5$ & $+35.6$ \\
$-1$ & 12.8 & 1.5603 & 1.5242 & 1.5328 & 1.5300 & $+30.3$ & $+36.1$ & $+29.9$ \\
$0$ & 0 & 1.5607 & 1.5366 & 1.5366 & 1.5366 & $+24.1$ & $+24.1$ & $+24.1$ \\
$+1$ & $-12.2$ & 1.5610 & 1.5494 & 1.5403 & 1.5433 & $+17.7$ & $+11.6$ & $+18.1$ \\
$+2$ & $-23.6$ & 1.5611 & 1.5627 & 1.5438 & 1.5501 & $+11.0$ & $-1.6$ & $+11.9$ \\
$+3$ & $-34.4$ & 1.5612 & 1.5766 & 1.5471 & 1.5569 & $+4.2$ & $-15.4$ & $+5.5$ \\
\bottomrule
\end{tabular}
\end{table}

\textbf{Coherency in the observed orientation.} The HD periods along $[1\bar{2}10]$ and $[0001]$
were locked to the graphite $a$ and $\sqrt3a$ periods (2.4565 and 4.2548~\AA{} at 0~GPa; 2.4242
and 4.1988~\AA{} at 20~GPa, r$^2$SCAN+rVV10). This strains the HD $a$ axis by $-1.83\%$ and its
$c$ axis by $+2.09\%$ (ambient), or by $-3.13$ and $+0.74\%$ (20~GPa). With the third axis relaxed (a
three-point energy scan; the stress along that axis crosses zero inside the scan) the asymmetry
is $+66.1$~m\AA{} at ambient ($\OB=1.5947$, $\langle\OA\rangle=1.5285$~\AA) and $+52.1$~m\AA{}
against the 20-GPa graphite lattice ($\OB=1.5712$, $\langle\OA\rangle=1.5191$~\AA). The confining
stresses are $\sigma_{xx}=+24$, $\sigma_{zz}=-26$~GPa and $+42$, $-10$~GPa respectively.

\textbf{E-mode splitting as a stress diagnostic.} Under uniaxial in-plane strain the two E
modes lose their degeneracy (Table~\ref{tab:esplit}). Each pair splits by $\approx$19--20~\icm{}
per per cent of strain, roughly 2~\icm{} per GPa of uniaxial stress, while the $A_{1g}$ frequency
moves by only $+4.5$ ($-3.4$)~\icm{} per per cent of compression (tension). A resolved E-mode
splitting therefore signals residual uniaxial in-plane stress. Lai \textit{et al.}\ fit a single
$E_{1g}$ component, which indicates that no splitting was resolved at the component width (a
splitting of 10~\icm{} would correspond to 0.5\% of strain). The published data do not fix the
splitting that would have been detectable, so we quote this as an indication rather than a
bound. The inversion remains accurate under such strain, and these strained structures played no
part in the calibration of the map (Section~S2). Reading the mean of the split $E_{1g}$ pair
against $A_{1g}$ through 0.41~m\AA{} per \icm{} gives $34.2$, $29.5$, $18.3$ and $13.2$~m\AA{} at
$-2$, $-1$, $+1$ and $+2\%$, against the directly computed mean asymmetries $36.4$, $30.3$, $17.7$
and $11.0$~m\AA{} (largest deviation 2.2~m\AA). With the measured-cell structure of Section~S1
(direct 24.0, inferred 24.2~m\AA) these form main-text Fig.~3c. The comparison separates two
things: the true asymmetry changes with strain by up to 13~m\AA{} over this range, while the
error of applying the equilibrium calibration to the strained structure stays within 2.2~m\AA.

\begin{table}[h]\centering\footnotesize
\setlength{\tabcolsep}{6pt}
\caption{$\Gamma$ frequencies (\icm) of the Raman-active manifold under clamped uniaxial strain
along HD$[10\bar{1}0]$ ($3\times3\times2$ finite displacements; $D_{2h}$ labels in the deposit).}
\label{tab:esplit}
\begin{tabular}{rccccc}
\toprule
$\varepsilon$ (\%) & $E_{2g}$ pair & $A_{1g}$ & $E_{1g}$ pair & $E_{2g}$ splitting & $E_{1g}$ splitting \\
\midrule
$-2$ & 1{,}228 / 1{,}265 & 1{,}312 & 1{,}345 / 1{,}384 & 37 & 39 \\
$-1$ & 1{,}219 / 1{,}239 & 1{,}306 & 1{,}337 / 1{,}357 & 19 & 20 \\
$0$ & 1{,}211 & 1{,}301 & 1{,}329 & 0 & 0 \\
$+1$ & 1{,}182 / 1{,}202 & 1{,}298 & 1{,}302 / 1{,}322 & 20 & 19 \\
$+2$ & 1{,}153 / 1{,}194 & 1{,}295 & 1{,}278 / 1{,}314 & 41 & 36 \\
\bottomrule
\end{tabular}
\end{table}

\FloatBarrier
\section{Graphite/diamond interface model}\label{sec:interface}

{\sloppy Each cell contains six HD $(10\bar{1}0)$ prismatic planes and six graphene layers (48 atoms, four
per plane) stacked along the common axis HD$[10\bar{1}0]\parallel$G$[0001]$, with
HD$[1\bar{2}10]\parallel$G$[1\bar{2}10]$ and HD$[0001]\parallel$G$[10\bar{1}0]$. The lateral cell
is fixed at either the HD lattice (graphite strained by $+1.9$ and $-2.0\%$) or the graphite
lattice (HD strained by $-1.8$ and $+2.1\%$). Three starting geometries were relaxed for each
lateral cell: the first graphene layer 1.8~\AA{} or 3.3~\AA{} above the outermost diamond atoms,
and a sheet placed at the gallery spacing (1.445~\AA). In the last, two of the sheet's atoms sit
at bond distance from the two dangling-bond atoms of the diamond surface and the other two are
lifted by 0.4~\AA{} (boat buckling). Stage~1 (600~eV, $13\times7\times1$ mesh) was followed by stage~2 (800~eV,
$19\times11\times1$) with the cell length corrected to zero stress along the stacking axis;
residual forces are 0.005--0.012~eV\,\AA$^{-1}$ and $|\sigma_{zz}|\le0.4$~GPa.\par}

\begin{figure}[h]\centering
\includegraphics[width=120mm]{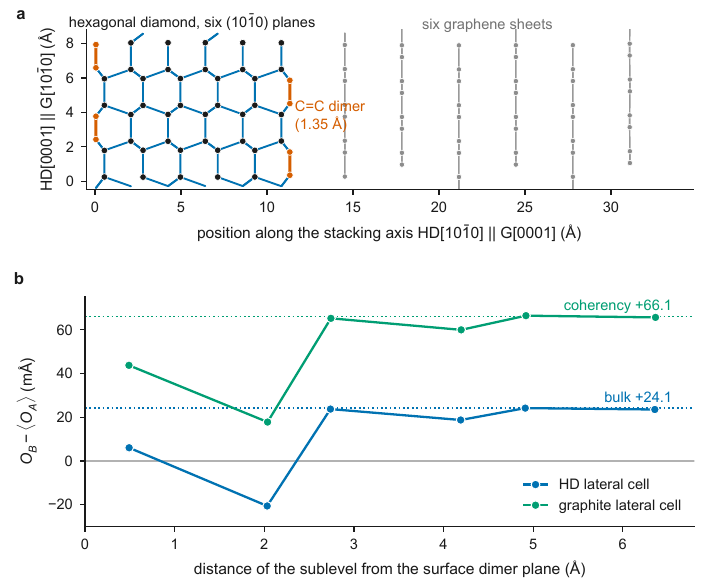}
\caption{The relaxed graphite/diamond interface. (a)~The HD-lateral cell started at van der Waals
distance (800~eV), viewed along HD$[1\bar{2}10]\parallel$G$[1\bar{2}10]$: six puckered
$(10\bar{1}0)$ planes of diamond (bonds in blue, the surface C=C dimers in vermilion) and six
graphene sheets (grey), with no bond across the 3.2~\AA{} gap. (b)~$\OB-\langle\OA\rangle$ per
sublevel against the distance from the surface dimer plane for the HD-lateral and
graphite-lateral cells (Table~\ref{tab:iface}); dotted lines mark the bulk value and the
coherency value of Section~S10.}
\label{fig:iface}
\end{figure}

All six relaxations converge to the same interface. The outermost pair of the diamond surface,
an axial pair, forms a carbon--carbon dimer of 1.349--1.358~\AA, and its two in-sheet basal bonds
contract to 1.478--1.489~\AA. The gallery-spanning basal bond of the next sublevel stretches to
1.63--1.64~\AA, and the nearest graphene sheet settles 3.20--3.25~\AA{} above the dimer plane with
no covalent bond to the diamond (Fig.~\ref{fig:iface}a). The starts differ in where the diamond
ends. A sheet started at van der Waals distance does not bond. The pre-bonded sheet stays bonded:
its two bonds to the surface relax from 1.54--1.60 to 1.63~\AA{} and its two lifted atoms pair
into the 1.35~\AA{} dimer. It thus becomes the reconstructed surface plane of a seven-plane
diamond block facing five free sheets. Table~\ref{tab:iface} gives the bond profile of
the 800-eV pre-bonded cells sublevel by sublevel, and Fig.~\ref{fig:iface}b plots it; the two
starts at each lateral cell agree to 0.5~m\AA. The perturbation is largest in the two
outermost sublevels. The third and fourth sublevels lie within 6~m\AA{} of the interior value, and
the two innermost sublevels recover it ($+24$~m\AA{} at the HD lattice; $+66$~m\AA{} at the
graphite lattice, the coherency value of Section~S10). The slabs are too thin to fix a decay length beyond
these six sublevels, and the models do not give the interface fraction of any sample, so they
bound the extent of the structural perturbation, not its spectroscopic weight.

\begin{table}[h]\centering\footnotesize
\setlength{\tabcolsep}{4.2pt}
\caption{Bond profile across the relaxed graphite/diamond interface (800~eV, pre-bonded start;
distances from the reconstructed surface dimer plane). The dimer pair itself (distance 0,
three-coordinated, its former axial bond being the 1.35~\AA{} dimer) is not listed.}
\label{tab:iface}
\begin{tabular}{ccccccc}
\toprule
 & \multicolumn{3}{c}{HD lateral cell} & \multicolumn{3}{c}{graphite lateral cell} \\
\cmidrule(lr){2-4}\cmidrule(lr){5-7}
sublevel distance (\AA) & \OB{} (\AA) & $\langle\OA\rangle$ (\AA) & $\OB-\langle\OA\rangle$ (m\AA) & \OB{} (\AA) & $\langle\OA\rangle$ (\AA) & $\OB-\langle\OA\rangle$ (m\AA) \\
\midrule
0.49 & 1.5424 & 1.5364 & $+6.0$ & 1.5763 & 1.5325 & $+43.8$ \\
2.03 & 1.5448 & 1.5654 & $-20.6$ & 1.5767 & 1.5588 & $+17.8$ \\
2.74 & 1.5599 & 1.5362 & $+23.7$ & 1.5932 & 1.5279 & $+65.3$ \\
4.19 & 1.5578 & 1.5391 & $+18.7$ & 1.5910 & 1.5310 & $+60.0$ \\
4.91 & 1.5608 & 1.5367 & $+24.2$ & 1.5948 & 1.5282 & $+66.5$ \\
6.36 (centre) & 1.5604 & 1.5368 & $+23.6$ & 1.5942 & 1.5285 & $+65.7$ \\
\bottomrule
\end{tabular}
\end{table}

\FloatBarrier
\section{Bond force constants}\label{sec:fc}

The longitudinal (stretch) force constant of each bonded pair is $k=-\hat{u}^{\rm T}\Phi_{ij}\hat{u}$,
with $\Phi_{ij}$ the symmetrized interatomic force-constant block from the finite-displacement
calculation and $\hat{u}$ the unit bond vector. It is 14.21~eV\,\AA$^{-2}$ for the basal bonds and
12.11~eV\,\AA$^{-2}$ for the axial bond of relaxed HD ($3\times3\times2$ supercell), against
13.94~eV\,\AA$^{-2}$ for the bond of cubic diamond ($4\times4\times4$ supercell). The bending
constants (the two eigenvalues of $-\Phi_{ij}$ transverse to the bond) are 7.1--7.4~eV\,\AA$^{-2}$
for all three. The axial bond is therefore longer, weaker by integrated crystal orbital Hamilton
population ($-9.35$ versus $-9.77$~eV) and 15\% softer than the basal bonds (which are 17\%
stiffer than it), a larger contrast than the 4.5\% of the bond populations. The softer axial bond
is the one whose stretch the $A_{1g}$ mode gauges.

\FloatBarrier
\section{Confinement and thermal bounds on the inversion}\label{sec:bounds}

\textbf{Phonon confinement.} In the Richter--Campbell model a crystallite of size $L$ relaxes
wavevector conservation with a Gaussian weight $|C(q)|^2=\exp(-q^2L^2/4)$, and the first-order
line becomes $I(\omega)\propto\int d^3q\,|C(q)|^2/[(\omega-\omega_b(q))^2+(\Gamma/2)^2]$ for each
branch $b$. The branch dispersions come from the archived force constants, with branches followed from
$\Gamma$ by eigenvector overlap and $\Gamma=8$~\icm. The integral is evaluated on a spherical
wavevector grid of 30 radial $\times$ 12 polar $\times$ 24 azimuthal points to
$|q|=0.25$~\AA$^{-1}$, and peak positions are read on a 0.1-\icm{} frequency grid. The peak shifts
are $-0.4$, $-0.5$ and $-0.6$~\icm{} for $E_{2g}$, $A_{1g}$ and $E_{1g}$ at $L=3$~nm,
$-0.2$~\icm{} for all three at 5~nm, and below 0.1~\icm{} at 10~nm and beyond. Doubling the grid
in every direction and halving the frequency step changes each shift by at most 0.05~\icm. The
differential shift between $E_{1g}$ and $A_{1g}$, the quantity that enters the inversion, is at
most 0.15~\icm{} at every size on either grid, i.e.\ $\le0.06$~m\AA. The model describes relaxed
wavevector selection in a crystallite of the specified structure. It does not describe the
altered force constants of reconstructed boundaries or of stacking-disordered material. Their
structural perturbation is confined to a few planes (Section~S11, main-text Fig.~2d), and their
effect on a fitted component position enters the analysis through the component-fit uncertainty
of Section~S1. The bound is quoted for lamellae down to 3~nm and does not rely on the long-range
perfection of the samples.

\textbf{Temperature.} The mode Gr\"uneisen parameters from $\Gamma$ phonons at the cells relaxed at
0, 5 and 10~GPa ($V=22.60$, 22.36 and 22.13~\AA$^3$ per four atoms; $E_{2g}$ 1{,}211.0 / 1{,}225.7
/ 1{,}240.4, $A_{1g}$ 1{,}301.2 / 1{,}315.5 / 1{,}329.4, $E_{1g}$ 1{,}329.2 / 1{,}343.0 /
1{,}356.5~\icm) are $\gamma=1.14$, 1.02 and 0.96. With the volumetric expansion of diamond
between 0 and 300~K, $\Delta V/V\approx2.8\times10^{-4}$, the quasi-harmonic shifts are $-0.39$,
$-0.37$ and $-0.36$~\icm, and their differential $E_{1g}-A_{1g}$ part is 0.01~\icm, i.e.\
$\le0.01$~m\AA{} on the asymmetry: the calculated thermal-expansion contribution nearly cancels in
the splitting. The explicit phonon--phonon self-energy shift amounts
to about $-1$ to $-2$~\icm{} for the diamond line between 0 and 300~K and includes a zero-point
term relative to the static harmonic reference [S8]. It has not been computed mode by mode here,
and the calculations above do not bound its differential part. The budget of Section~S1 therefore carries an allowance
rather than a bound. A differential shift as large as the whole 300-K shift of the diamond line,
2~\icm, would move the asymmetry by 0.8~m\AA{} and leaves the rounded interval at $\pm3$~m\AA. For
scale, differential shifts of 5 and 10~\icm{} would move it by 2.1 and 4.1~m\AA{} (sensitivity
scenarios, not established effects).

\FloatBarrier
\section*{Supplementary references}

\begin{list}{}{\setlength{\leftmargin}{2.2em}\setlength{\labelwidth}{2em}\setlength{\itemsep}{0pt}}
\item[{[S1]}] Cromer, D. T. \& Mann, J. B. X-ray scattering factors computed from numerical
Hartree--Fock wave functions. \textit{Acta Crystallographica Section A} \textbf{24}, 321--324
(1968).
\item[{[S2]}] Yang, L. \textit{et al.} Synthesis of bulk hexagonal diamond. \textit{Nature}
\textbf{644}, 370--375 (2025).
\item[{[S3]}] Lai, S. \textit{et al.} Bulk hexagonal diamond. \textit{Nature} \textbf{651},
621--625 (2026).
\item[{[S4]}] Chen, D. \textit{et al.} General approach for synthesizing hexagonal diamond by
heating post-graphite phases. \textit{Nature Materials} \textbf{24}, 513--518 (2025).
\item[{[S5]}] Goryainov, S. V., Likhacheva, A. Yu. \& Ovsyuk, N. N. Raman scattering in
lonsdaleite. \textit{Journal of Experimental and Theoretical Physics} \textbf{127}, 20--24
(2018).
\item[{[S6]}] Yoshiasa, A., Murai, Y., Ohtaka, O. \& Katsura, T. Detailed structures of
hexagonal diamond (lonsdaleite) and wurtzite-type BN. \textit{Japanese Journal of Applied
Physics} \textbf{42}, 1694--1704 (2003).
\item[{[S7]}] Dollase, W. A. Correction of intensities for preferred orientation in powder
diffractometry: application of the March model. \textit{Journal of Applied Crystallography}
\textbf{19}, 267--272 (1986).
\item[{[S8]}] Lang, G. \textit{et al.} Anharmonic line shift and linewidth of the Raman mode in
covalent semiconductors. \textit{Physical Review B} \textbf{59}, 6182 (1999).
\end{list}

Experimental values quoted in this Supplementary Information are from refs.~[S2] (Yang
\textit{et al.}), [S3] (Lai \textit{et al.}), [S4] (Chen \textit{et al.}), [S5] (Goryainov
\textit{et al.}) and [S6] (Yoshiasa \textit{et al.}); the texture proxy follows ref.~[S7] and the
anharmonic shift of the diamond line ref.~[S8]. The
density-functional and phonon methodology references are given in the main-text reference list.

\end{document}